\documentclass[acmsmall]{acmart}
\usepackage{soul}
\usepackage{epstopdf}
\usepackage{hyperref}
\usepackage{multirow}
\usepackage{listings}
\usepackage{fancybox}
\usepackage{graphicx}
\usepackage{subcaption}
\usepackage{paralist}
\usepackage{ragged2e}
\usepackage{enumitem}
\usepackage{booktabs}
\usepackage{longtable}
\usepackage{supertabular}

\usepackage{ulem,xcolor,tikz}

\usepackage[figuresright]{rotating}

\begin{document}

\title{A Roadmap for Software Testing in Open-Collaborative and
AI-Powered Era}


\author{Qing Wang}
\email{wq@iscas.ac.cn}
\authornote{Also With State Key Laboratory of Intelligent Game, Beijing, China; \\ University of Chinese Academy of Sciences, Beijing, China;\\ }
\authornote{Corresponding authors}
\affiliation{
  \institution{Institute of Software Chinese Academy of Sciences}
  \city{Beijing}
  \country{China}
}

\author{Junjie Wang}
\email{junjie@iscas.ac.cn}
\authornotemark[1]
\authornotemark[2]
\affiliation{
  \institution{Institute of Software Chinese Academy of Sciences}
  \city{Beijing}
  \country{China}
}

\author{Mingyang Li}
\email{mingyang2017@iscas.ac.cn}
\authornotemark[1]
\affiliation{
  \institution{Institute of Software Chinese Academy of Sciences}
  \city{Beijing}
  \country{China}
}

\author{Yawen Wang}
\email{yawen2018@iscas.ac.cn}
\authornotemark[1]
\affiliation{
  \institution{Institute of Software Chinese Academy of Sciences}
  \city{Beijing}
  \country{China}
}

\author{Zhe Liu}
\email{liuzhe2020@iscas.ac.cn}
\authornotemark[1]
\affiliation{
  \institution{Institute of Software Chinese Academy of Sciences}
  \city{Beijing}
  \country{China}
}

\renewcommand{\shortauthors}{Wang, et al.}

\begin{abstract}
Internet technology has given rise to an open-collaborative software development paradigm, necessitating the open-collaborative schema to software testing.
It enables diverse and globally distributed contributions, but also presents significant challenges to efficient testing processes, coordination among personnel, and management of testing artifacts.
At the same time, advancements in artificial intelligence (AI) have enhanced testing capabilities and enabling automation, while also introducing new testing needs and unique challenges for AI-based systems.
In this context, this paper explores software testing in the open-collaborative and AI-powered era, focusing on the interrelated dimensions of process, personnel, and technology.
Among them, process involves managing testing workflows and artifacts to improve efficiency, personnel emphasizes the role of individuals in ensuring testing quality through collaboration and contributions, while technology refers to AI methods that enhance testing capabilities and address challenges in AI-based systems.
Furthermore, we delve into the challenges and opportunities arising from emerging technologies such as large language models (LLMs) and the AI model-centric development paradigm.
\end{abstract}




\maketitle

\section{Introduction}
\label{sec_introduction}

Open collaboration is a hallmark of the Internet era in the past ten to twenty years, where geographical distances cease to be a barrier. Platforms like social media networks and online forums have emerged as global hubs for communication, enabling individuals worldwide to exchange ideas, share experiences, and engage in discussions \cite{gruzd2013enabling}. 
Similarly, in the realm of software development, this has led to the rise of the open collaborative development paradigm \cite{crowston2008free}. 
Platforms like GitHub have flourished, providing developers from diverse backgrounds with the opportunity to contribute code, discuss technical matters, and address software issues, regardless of their geographic location \cite{booch2003collaborative,sengupta2006research}.
In such development environments, software testing and quality assurance faces both new challenges and opportunities. 
Unlike the closed-environment testing paradigm with fixed contributors and predefined setups, open-environment testing faces the challenges of distributed collaboration, diverse contributors, and rapid iterations, often resulting in uneven outcomes.
This variability necessitates not only a focus on testing techniques but also on the testing process itself, such as improving coordination of testing activities to handle variability and streamlining the management of testing artifacts to achieve better cost-effectiveness. Additionally, attention must also be given to the personnel involved, providing them with the support needed to enhance their efficiency.

Another significant shift is the rise of artificial intelligence (AI) technologies. 
From the early days of machine learning and deep learning to the advent of pre-trained large language models (LLMs), these advancements offer effective solutions to address the challenges inherent in open-collaborative testing. 

There exist notable roadmaps that have significantly contributed to the understanding and advancement of software testing practices. One such landmark roadmap is ``Software Testing: A Research Travelogue (2000–2014),'' spearheaded by Professor Alessandro Orso and Gregg Rothermel \cite{Orso2014softwareTesting}. 
This seminal work primarily explores software testing techniques and methodologies, offering valuable insights into the evolution of testing technologies over the past decade and a half. 

However, a decade has passed since then, and with the constant emergence of new technologies, there is a pressing need for a new roadmap to summarize and project future research in software testing. 
Moreover, we aim to adopt a broader perspective, recognizing that the factors influencing software testing extend well beyond the techniques alone.
Therefore, a comprehensive overview and roadmap for software testing in the open-collaborative and AI-powered era requires a holistic understanding that encompasses not only testing methodologies but also broader contextual factors. 

\begin{figure*}[t!]
\centering
\includegraphics[width=\linewidth]{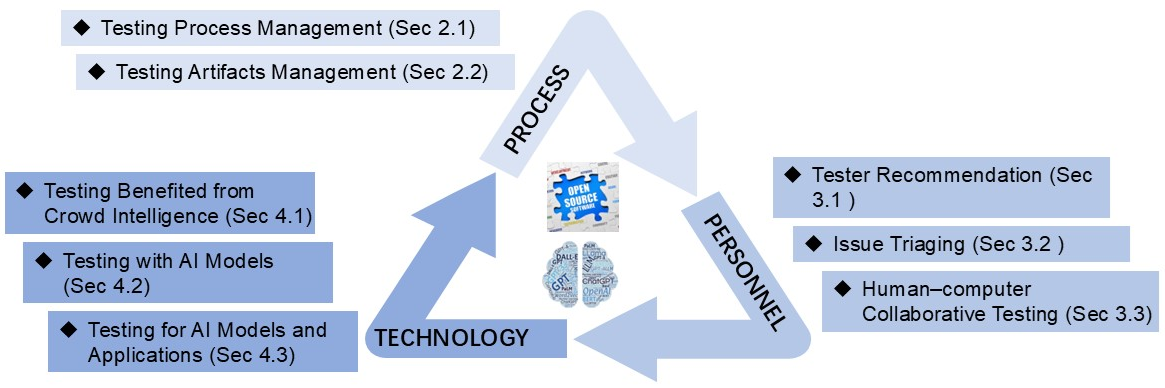}
\caption{Overview about software testing in open-collaborative and AI-powered era}
\label{fig:overview}
\vspace{-0.1in}
\end{figure*}

Taken in this sense, our exploration of software testing in the open-collaborative and AI-powered era is guided by the recognition that three key dimensions—process, personnel, and technology—form a triad of fundamental factors influencing software testing practices. 
These dimensions represent orthogonal facets that collectively shape the landscape of software testing, as visually demonstrated in Figure \ref{fig:overview}.
Among them, the \textbf{process} dimension focuses on methodologies to manage dynamic, distributed contributions efficiently and ensure timely, comprehensive testing coverage.
The \textbf{personnel} dimension highlights the critical role of human contributions in identifying and resolving issues to maintain testing quality.
The \textbf{technology} dimension emphasizes the impact of AI advancements in enhancing testing capabilities and addressing the unique challenges of AI-driven systems.
By examining these three dimensions—process, personnel, and technology—we gain a comprehensive understanding of the intricate landscape of software testing within open-collaborative environments.


Additionally, we also outline future trends.
As emerging technologies like LLMs continue to evolve, the landscape of software testing research is expected to expand to encompass various aspects. This may include leveraging LLMs for enhancing testing practices, exploring testing methodologies tailored for LLMs and related applications, ensuring quality assurance for auto-generated code produced by AI models, and refining collaboration strategies within AI model-centric paradigms. 
These areas represent promising avenues for future research and innovation in the field of software testing under open collaborative and AI-powered era.

To curate the relevant works, we adopt an informal methodology informed by the authors' expertise in software testing and AI. 
The selection prioritized survey papers and high-impact studies from top-ranked venues in software engineering 
as well as AI related venues.
This allowed us to quickly incorporate emerging insights into our discussion while focusing on the core challenges and opportunities related to software testing in the AI era.

The paper is structured as follows: Section 2 to 4 provide an overview of software testing researches from the perspectives of process, personnel, and technology, respectively. 
Section 5 explores the opportunities and challenges in this field, while Section 6 concludes the paper.



\section{Process Related}
\label{sec_process_related}

In open-collaborative environments, the process dimension of software testing involves practices and methodologies that enable efficient and effective testing in dynamic, distributed development settings, alongside the streamlined management of testing artifacts.

\subsection{Testing Process Management}

In modern software development, Continuous Integration (CI) has become a cornerstone practice for maintaining code quality and fostering collaboration. 
By frequently integrating code into a shared version control repository, CI ensures that each change is automatically validated through automated build and testing pipelines. 
A critical aspect of this process is regression testing, which serves as a key manifestation of the testing process by verifying that new changes do not introduce defects into existing functionality.
While regression testing is invaluable for maintaining software quality, executing a full suite of tests can be highly resource-intensive and time-consuming, often requiring hours or even days to complete. This delay can hinder the CI cycle and prevent timely feedback for developers \cite{yaraghi2022scalable}. 
To mitigate these challenges, techniques such as test case prioritization and test case selection have been developed to optimize the testing process, by selecting and prioritizing test cases in order to provide early feedback to developers.
We introduce the Test Case Prioritization (TCP) techniques which share similarities with selection techniques.


Earlier attempts start with heuristics-based techniques, e.g., Elbaum et al. \cite{Elbaum2014techniques} prioritized tests based on whether they have not been executed for long or have failed in the recent commits, and Haghighatkhah et al. \cite{Haghighatkhah2018Test} scheduled tests based on the combination of their previous execution results and similarity.
Later, other researchers harness the power of machine learning by using a large amount of historical data in CI, and propose numerous machine learning based TCP
techniques which have been demonstrated to be promising.
They build neural models to predict the optimal sequence of tests instead of human-defined strategies.
In particular, these machine learning based techniques can be categorized
into supervised learning-based \cite{Busjaeger2016Learning,Bertolino2020Learning,Sharif2021DeepOrder,yaraghi2022scalable}, and reinforcement learning-based techniques \cite{Spieker2017Reinforcement,Bagherzadeh2022Reinforcement,Lima2022Multi-Armed}.
The latter ones would continuously adjust its prioritization strategy, i.e., it is first tested (i.e., prioritizing the test suite) and then
trained based on the prioritization feedback.

Apart from that, in open-collaborative environments, crowdsourced testing (also typically abbreviated as crowdtesting) has become a key strategy for enhancing software quality by leveraging contributions from a large, diverse group of external testers. 
Crowdtesting is hard to manage in nature. Given the  unpredictability of distributed crowdtesting processes, it is difficult to estimate (a) remaining number of bugs yet to be detected or (b) required cost to find those bugs. 
Experience-based decisions may result in ineffective crowdtesting processes, e.g., there is an average of 32\% wasteful spending in current crowdtesting practices.
Wang et. al. \cite{wang2019isense} explored automated decision support
to effectively manage crowdtesting processes.



\subsection{Testing Artifacts Management}



In open-collaborative environments, the testers—often from different backgrounds and geographical locations—can provide valuable feedback by identifying issues that might not be caught through traditional testing methods. However, the volume of contributions from such a wide range of sources presents unique challenges.
The influx of issue reports, bug findings, and feedback (all of them can be treated as testing artifacts) can lead to information overload, where valuable insights are buried under redundant or conflicting contributions. This makes it difficult to prioritize and filter out irrelevant data, which can undermine the efficiency of the testing process.
Effective information filtering techniques are therefore necessary to improve the signal-to-noise ratio and streamline the testing process.
Duplicate detection is the most-commonly employed technique, which aims at identifying and eliminating redundant or duplicate information, such as duplicate issue reports or discussions on similar topics. 
For example, Nguyen et. al. \cite{Nguyen2012duplicate} applied information retrieval techniques for duplicate detection by computing the textual similarity between two reports. 
Sun et. al. \cite{sun2010adiscriminative} designed a set of features for measuring the reports’ similarity in terms of textual descriptions and attributes, and employed machine learning techniques for duplicate detection. 
Yang et. al. \cite{yang2016combining} modeled the semantic similarity of reports using deep learning techniques for duplicate detection.
In addition, Huang et. al. \cite{huang2020questfor} and Zhang et. al. \cite{zhang2023duplicate} respectively conducted  experimental evaluations of the commonly-used approaches for duplicate detection.


\section{Personnel Related}
\label{sec_people_related}

In open-collaborative software testing, the involvement of individuals plays a crucial role in the success of the testing process. 
Human factors influence the quality of testing, as contributions and collaborations from testers, developers, and stakeholders are vital for identifying critical issues, quickly fix the issues, and ensuring comprehensive coverage. 


\subsection{Tester Recommendation}

Unlike typical software development projects, testing tasks typically require a group of testers, ideally with diverse backgrounds, to achieve the goal of diversified testing and covering different areas of the software. 
Therefore, different from the worker recommendation for general crowdsourcing tasks, the tester recommendation for crowdtesting tasks tend to take into account the diversity of the recommended workers. 
For example, Wang et al. \cite{wang2021characterizing} propose a multi-objective crowd tester recommendation approach, which aims at recommending crowd tester by maximizing the bug detection probability of testers, the relevance with the test task, the diversity of testers, and minimizing the test cost. 
Additionally, previous studies on worker recommendation mainly focus on one-time recommendations with respect to the initial context at the beginning of a new task. 
However, for crowdtesting, a typical task can last from 3 days to 2 weeks, during which crowd testers can freely conduct the testing and submit reports. 
Taking this into account, Wang et al. \cite{wang2020contextaware} point out the need for accelerating crowdtesting by recommending appropriate testers in a dynamic manner, and they propose a context-aware in-process crowd testers recommendation approach, to detect more bugs earlier and potentially shorten the testing period.

\subsection{Issue Triaging}
\label{sec_people_issue_reports}

Various issues appear during software testing, and issue fixing is a time-consuming and costly task.
Once an issue report is received, assigning it to a suitable developer within a short time interval can reduce the time and cost of the issue fixing process. This assignment process is known as issue triaging.
Issue triaging is a time-consuming process since often a large number of developers are involved in software testing.
To aid in finding appropriate developers, earlier practice adopted techniques as machine learning, graph analysis, fuzzy set, and topic modeling.
For example, Anvik et al. \cite{anvik2006should} utilized machine learning methods to solve it. 
Jeong et al. \cite{jeong2009improving} proposed to use a bug tossing graph to improve issue triaging prediction accuracy. 
Tamrawi et al. \cite{tamrawi2011fuzzy} proposed a method called Bugzie, which uses a fuzzy set and cache-based approach to increase the accuracy of issue triaging.
Xia et al. \cite{xia2017improving} proposed a specialized topic modeling algorithm named multi-feature topic model for issue triaging.
Later studies used deep learning for bug triaging, e.g., Lee et al. \cite{lee2017applying} applied word embedding to train a CNN based classifier.
Dipongkor et al. \cite{Dipongkor3023comparative} conducted the experimental evaluation about fine-tuning the transformer-based language models for this task.




\subsection{Human–computer Collaborative Testing}



There are studies that analyze incorporating automation technologies to assist manual testing, exemplifying human-computer interaction in the software testing process. 
For instance, Liu et al. automatically trace testers' actions and use explicit visual annotations to guide or remind them of unexplored areas, helping testers avoid missing functionalities or repeating steps \cite{liu2022guided}.
Similarly, Chen et al. investigate human collaboration in crowd testing scenarios. 
They utilize interactive event-flow graphs to track and aggregate each tester's interactions into a single directed graph, which visualizes the test cases already explored. 
Crowd testers can interact with these graphs to discover new navigation paths and improve test coverage \cite{chen2020improving}.

\section{Technology Related}
\label{sec_technology_related}

Advancements in AI technology have significantly enhanced testing capabilities, introducing intelligent automation and improving efficiency in open-collaborative environments. At the same time, the rise of AI-driven applications and systems has created new demands for testing, necessitating specialized techniques to validate the functionality, robustness, and fairness of AI models and applications. These dual developments highlight the evolving role of technology in both strengthening traditional testing processes and addressing the unique challenges posed by AI-centric systems.

\subsection{Testing Benefited from Crowd Intelligence}
\label{sec_technology_crowd_intelligence}

In the context of open collaborative software development, a vast amount of data contributed by developers with diverse backgrounds is aggregated. 
This data encapsulates rich knowledge about software quality assurance, and  can be harnessed to empower the testing techniques. 
For example, Mao et al. \cite{mao2017crowdintelligence} extracted the test scripts from crowdbased testing to automatically infer the reusable high-level event sequences for enhancing the  automated mobile testing.
Liu et al. \cite{liu2021owleyes} collected a large number of GUI screenshots with UI display issues from the crowdsourced testing platform, and used them to train a visual understanding model for automatically detecting the GUIs with display issues and locating the buggy region.
Wei et al. \cite{wei2022freelunch} mined the code/models from open source to obtain the code snippets from the deep learning library documentation, library developer tests, and deep learning models in the wild, then leveraged this  information to perform fuzz testing for deep learning libraries.

\subsection{Testing with AI Models}

\begin{figure*}[t!]
\centering
\includegraphics[width=\linewidth]{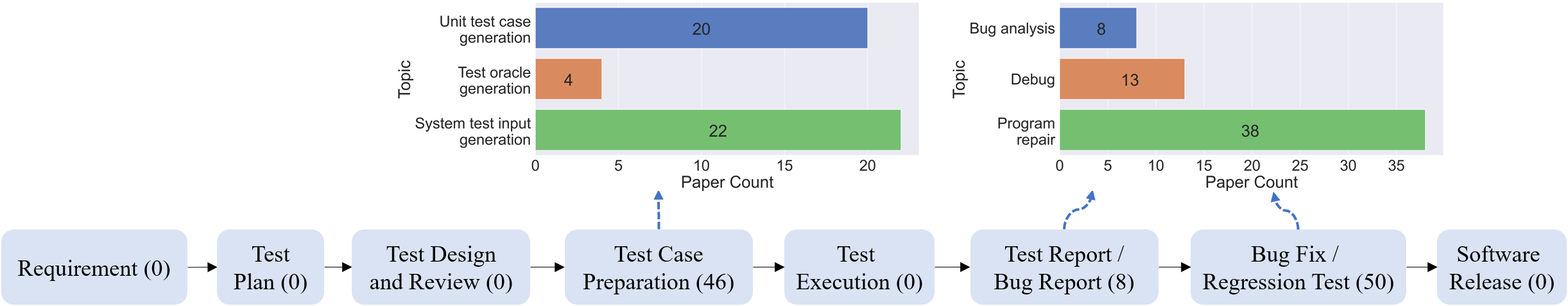}
\caption{Distribution of testing tasks with LLMs \cite{wang2023software}}
\label{fig:LLMforTest}
\vspace{-0.1in}
\end{figure*}

Since the introduction of the concept of training ``Deep Neural Network'' by Geoffrey Hinton in 2006, deep learning has been increasingly adopted to develop cutting-edge tools for software testing research, thanks to its ability to enhance performance. 
A comprehensive survey \cite{yang2022survey} explores how deep learning is applied in testing areas such as test case generation, bug localization, and application testing.

In recent years, the pervasive advancements in LLMs have profoundly impacted various domains, including software testing. 
These models have been increasingly harnessed to bolster testing capabilities across different facets, ranging from enhancing test coverage in unit testing to diversifying test case generation in integration testing.
There is a relevant literature review titled ``Software Testing with Large Language Models: Survey, Landscape, and Vision'' \cite{wang2023software}, which provides a comprehensive overview of the utilization of LLMs in software testing. It analyzes 102 relevant studies that have employed LLMs for software testing, examining them from both the software testing and LLMs perspectives.
As demonstrated in Figure \ref{fig:LLMforTest},  LLMs are commonly used for test case preparation (including unit test case generation, test oracle generation, and system test input generation), program debugging, and bug repair. 
However, there is currently no practices for applying LLMs in the tasks of early testing life-cycle (such as test requirement, test plan, etc).

For unit test case generation, a majority of the earlier published studies adopt the pre-training or fine-tuning schema.
For example Alagarsamy et al. \cite{23a3testTestGeneration} first pre-trained the LLM with the
focal method and asserted statements to enable the LLM to
have a stronger foundation knowledge of assertions. 
They then fine-tuned the LLM for the test case generation task where
the objective is to learn the relationship between the focal
method and the corresponding test case.
By comparison, most later studies typically focus on how to design the prompt, to make the LLM better understanding the
context of the task. 
Yuan et al. \cite{64noUnitTest} performed an empirical study to evaluate ChatGPT’s capability of unit test generation with both a quantitative analysis and a user study in terms of correctness, sufficiency, readability, and usability. 
And results show that the generated tests still suffer from correctness issues, including diverse compilation errors and execution failures. 
To address this, they proposed an approach where ChatGPT was used to improve the quality of its own generated tests, using an initial test generator and an iterative test refiner. 
The iterative refiner followed a validate-and-fix approach, correcting compilation errors by prompting the LLM based on error messages and additional code context.

For system test input generation, it varies for specific types of software being tested. 
For mobile applications, test input generation requires a wide range of text inputs or operation combinations (e.g., clicking a button or long-pressing a list) to test the application's functionality and user interface \cite{26fillBlank,60GUITesting}. 
In contrast, for Deep Learning (DL) libraries, the test input consists of programs that cover diverse DL APIs \cite{13fuzzDeepLearningLibraries,19fuzzDeepLearningLibraries}.
For example, Liu et al. \cite{60GUITesting} formulates the test input generation of mobile GUI testing problem as a Q\&A task, which asks LLM to chat with the mobile apps by passing the GUI page information to the LLM.
The LLM then generates testing scripts (i.e., GUI operations) and executes them while receiving app feedback, iterating the process. 
The proposed GPTDroid also introduces a functionality-aware memory prompting mechanism that equips the LLM with the ability to retain testing knowledge of the whole process and conduct long-term functionality-based reasoning to guide exploration.
Deng et al. \cite{13fuzzDeepLearningLibraries} used both generative and infilling LLMs to generate and mutate valid/diverse input DL programs for fuzzing DL libraries.
The process starts with a generative LLM (CodeX) to generate a set of seed programs using target DL APIs. 
Next, part of the seed program is replaced with masked tokens, and an infilling LLM (InCoder) is used to fill in the masked tokens and generate new code.

\subsection{Testing for AI Models and Applications}

The rise of AI applications raises concerns about trustworthiness, particularly in safety-critical domains such as self-driving systems and medical treatments. 
Software testing plays a crucial role in detecting and addressing discrepancies between expected and actual behaviors in these applications.
However, testing AI systems presents unique challenges due to their statistical nature, evolving behavior, and the oracle problem. 

\begin{figure*}[t!]
\centering
\includegraphics[width=0.6\linewidth]{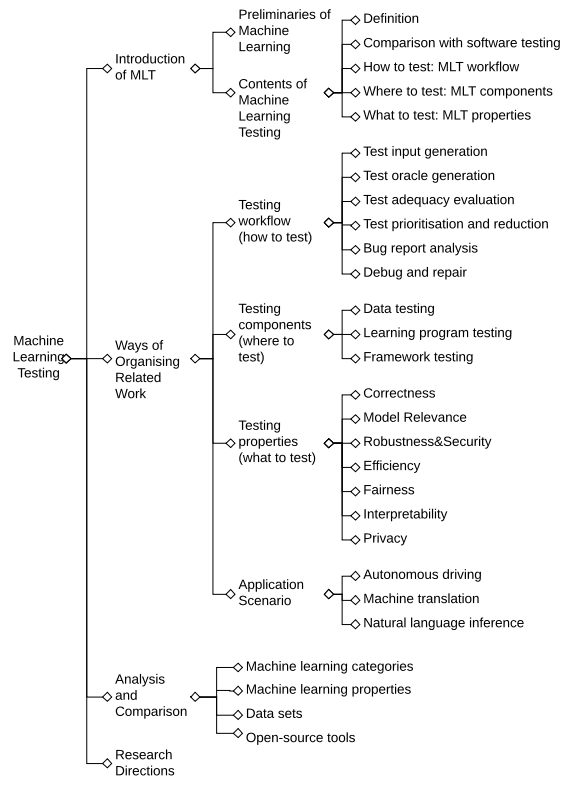}
\caption{Details of machine learning testing \cite{Zhang2022Machine}}
\label{fig:MLtest}
\vspace{-0.15in}
\end{figure*}

The notable systematic review titled ``Machine Learning Testing: Survey, Landscapes and Horizons'' \cite{Zhang2022Machine} presents an extensive examination of methodologies for assessing machine learning (including deep learning) systems.
It encompasses 144 papers that explore various aspects of testing properties (such as correctness, robustness, and fairness), testing components (including data, learning programs, and frameworks), workflows (encompassing test generation and evaluation), application scenarios (such as autonomous driving and machine translation), as shown in Figure \ref{fig:MLtest}.

Adversarial inputs represent a critical concept within AI testing, as they play a significant role in assessing the robustness of AI models. 
These inputs are deliberately perturbed based on the original inputs, often deviating from the typical data distribution encountered in real-world scenarios. 
By subjecting models to these carefully crafted inputs, testers can identify potential weaknesses and shortcomings of AI models.
For example, Zhou et al. \cite{zhou2020Deepbillboard} proposed DeepBillboard to generate realworld adversarial billboards that can trigger potential steering errors of autonomous driving systems.
Sun et al. \cite{sun2020automatic} automatically generate test inputs via mutating the words in translation inputs for testing machine translation systems. 
In order to generate translation pairs that ought to yield consistent translations, their approach conducts word replacement based on word embedding similarities. 

The test oracle problem is challenging, because many machine learning algorithms are probabilistic programs. 
Metamorphic relations are widely studied to tackle the oracle problem, and they are based on transformations of training or test data that are expected to yield unchanged or certain expected changes in the predictive output.
For example, Dwarakanath et al. \cite{DwarakanathASRB2018Identifying} applied metamorphic relations to image classifications with SVM and deep learning
systems. The changes on the data include changing the
feature or instance orders, linear scaling of the test features,
normalisation or scaling up the test data, or changing the
convolution operation order of the data.
Tian et al. \cite{tian2018deeptest} and Zhang et al. \cite{ZhangZZ0K2018Deeproad} stated that the
autonomous vehicle steering angle should not change significantly or stay the same for the transformed images under different weather conditions.

Test adequacy evaluation aims to discover whether the
existing tests have a good fault-revealing ability. It provides
an objective confidence measurement on testing activities.
In traditional software testing, code coverage measures the
degree to which the source code of a program is executed by
a test suite \cite{zhang2019predictive}. 
Unlike traditional software, code coverage is seldom a
demanding criterion for AI testing, since the decision logic
of an AI model is not written manually but rather it is
learned from training data. 
Pei et al. \cite{pei2017deepXplore} proposed the first coverage criterion, neuron coverage, particularly designed for deep learning testing. Neuron coverage is calculated as the ratio of the number of unique neurons activated by all test inputs and the total number of neurons in a DNN. In particular, a neuron is activated if its output value is larger than a user-specified threshold.
Following that, there are more fine-grained criteria, like k-multisection neuron coverage, neuron boundary coverage, and strong neuron activation coverage, etc.


\section{Challenges and Opportunities}

Despite the advancements and breakthroughs discussed in the preceding sections, open collaborative development environments and the rapid evolution of AI technologies/ecosystems continue to present numerous challenges, highlighting ongoing research opportunities and emerging trends in the coming years.


\subsection{Leveraging LLMs for Enhancing Testing}

Although software testing with LLMs has undergone significant growth in the past three years, it is still in its early stages of development, and numerous challenges and open questions need to be addressed.

\textbf{Tese case generation.}
Exploring diverse software behaviors while minimizing costs remains a critical challenge in software testing. Test case generation, in particular, poses significant obstacles in traditional automated testing due to the difficulty of producing test inputs with correct grammar and semantics, especially for complex or composite inputs. 
As a result, automated testing often struggles to achieve comprehensive coverage. 
LLMs offer promising potential with their exceptional ability to understand software context, yet current research has barely scratched the surface of their capabilities. 
For example, for unit test case generation, in SF110 dataset, the line coverage is merely 2\% \cite{66generatingUnitTests}.
For system test input generation, in terms of fuzzing DL libraries, the API coverage for TensorFlow is reported to be 66\% (2215/3316) \cite{13fuzzDeepLearningLibraries}.

As the initial wave of enthusiasm for LLMs has subsided, people have become increasingly aware of the limitations of LLMs and the strengths of program analysis and traditional testing techniques. 
Consequently, the integration of LLMs with traditional approaches has emerged as an important direction for future exploration.
One direction may utilize mutation testing together with the LLMs to generate more diversified outputs. 
For example, when fuzzing a DL library, instead of directly generating the code snippet with LLM, Deng et al. \cite{13fuzzDeepLearningLibraries} replace parts of the selected seed (code generated by LLM) with masked tokens using different mutation operators to produce masked inputs. 
They then leverage the LLM to perform code infilling to generate new code that replaces the masked tokens, which can significantly increase the diversity of the generated tests. 
Automatically generate test cases based on metamorphic relations to cover a wide range of inputs is also a promise avenue.
Combine LLM with other traditional technique is also a promising direction, e.g., Jiang et. al. \cite{jiang2024towards} conduct a systematic study on the LLMs and constraint-based tools for test input generation, and find that there are limitations for LLMs in specific scenarios such as sequential calculation, where constraint-based tools are
in a position of strength. By combining them together, there can be 1.4x to 2.3x improvement than the baselines.

Other potential research direction could involve utilizing testing-specific data to train or fine-tune a specialized LLM that is specifically designed to understand the nature of testing. By doing so, the LLM can inherently acknowledge the requirements of testing and autonomously generate diverse outputs.

\textbf{Test oracle problem. }
The oracle problem has been a longstanding challenge in various testing applications, e.g., testing machine learning systems \cite{zhang2022machineLearningTesting} and testing deep learning libraries \cite{13fuzzDeepLearningLibraries}. 
To alleviate the oracle problem to the overall testing activities, a common practice is to transform it into a more easily derived form, often by utilizing differential testing \cite{204SmtSolverValidation} or focusing on only identifying crash bugs \cite{60GUITesting}.

Exploring the use of LLMs to derive other types of test oracles represents an interesting and valuable research direction. 
Specifically, metamorphic testing is also widely used in software testing practices to help mitigate the oracle problem, yet in most cases, defining metamorphic relations relies on human ingenuity. 
Recent preliminary attempts have demonstrated the potential of leveraging LLMs to automatically discover and define these relations, offering a promising avenue for further investigation and application \cite{luu2023chatgptadvancesoftwaretesting}.
Another promising avenue is exploring the capability of LLMs to automatically generate test cases based on metamorphic relations, covering a wide range of inputs.
Apart from that, the advancement of multi-model LLMs like GPT-4 may open up possibilities for exploring their ability to detect bugs in software user interfaces and assist in deriving test oracles. 
By leveraging the image understanding and reasoning capabilities of these models, one can investigate their potential to automatically identify inconsistencies, errors, or usability issues in user interfaces, e.g., VisionDroid \cite{liu2024visiondrivenautomatedmobilegui} makes the first attempt towards this direction. 

\textbf{Real-world application of LLMs in software testing.}
Due to concerns regarding data privacy, when considering real-world practice, most software organizations tend to avoid using commercial LLMs and would prefer to adopt open-source ones with training or fine-tuning using organization-specific data. Furthermore, some companies also consider the current limitations in terms of computational power or pay close attention to energy consumption, they tend to
fine-tune medium-sized models. 
It might be quite challenging for these models to achieve similar performance to what existing papers have reported. 
Recent research has highlighted the importance of high-quality training data in improving the performance of models for code-related tasks.
A notable example in this area is the StarCoder2 model \cite{lozhkov2024starcoder2stackv2}, where the StarCoder2-15B significantly outperforms other models of similar size (e.g., CodeLlama-13B) and even matches or surpasses the performance of CodeLlama-34B. 
The superior performance of StarCoder2 is reported as attributing to the curation of high-quality open data sources, such as GitHub issues, pull requests, Kaggle datasets, Jupyter notebooks, and code documentation. Additionally, rigorous data preprocessing steps, including deduplication and the application of filters to eliminate low-quality code, have played a crucial role in enhancing the model's effectiveness.
However, building high-quality, organization-specific datasets for training or fine-tuning is a time-consuming and labor-intensive process. 
To address this challenge, automated techniques from the field of mining software repositories \cite{hassan2008road}—an area that has made significant strides over the past decade—can be employed. 
These techniques enable efficient extraction and analysis of key information, streamlining the dataset creation process.

In addition, exploring methodologies for better pre-training or fine-tuning LLMs with software-specific data is a promising direction. 
In recent times, the approach to LLM pre-training and fine-tuning has largely followed the ``more data, better results'' philosophy, with an emphasis on utilizing as much data as possible. 
However, the availability of open data is limited, and this path has nearly reached its limits. 
On the other hand, software-specific data differs from natural language data in that it contains more structural information, such as data flow and control flow. 
Therefore, enabling LLMs to better learn these software-specific structural elements may become a key focus in future research. 
Previous work on code representations has highlighted the benefits of incorporating data flow information, as demonstrated by Guo et al. [2021] in their GraphCodeBERT model.

\subsection{Testing for LLMs and LLM-centric Applications}

Since the emergence of LLMs, AI-enabled software applications have rapidly advanced. The quality of these applications now relies not only on functional correctness but also on the performance of the embedded or associated AI models. This shift has introduced unprecedented challenges to testing.

\textbf{Testing methodology specifically designed for LLMs.} There have been numerous research efforts on testing machine learning and deep learning models, however, in the context of general artificial intelligence, i.e., LLMs, there has been relatively less exploration in software-related conferences and journals. 
At AI-related conferences, much work has been done on benchmarking LLMs, such as evaluating their performance in task automation \cite{wang2024pandalmautomaticevaluationbenchmark}, instruction tuning \cite{shen2024taskbenchbenchmarkinglargelanguage}, and judge assistants \cite{chen2024mllmasajudgeassessingmultimodalllmasajudge}. 
These benchmarking efforts have significantly advanced the field, with some works already having hundreds of citations. 
From the perspective of software engineering, a systematic testing methodology specifically designed for LLMs is urgently needed.
It’s essential to extend testing beyond basic benchmarks and include functional, non-functional, and safety testing, especially considering the complex and unpredictable behaviors of LLMs when applied in real-world software applications.
Furthermore, with the emergence of multi-modal LLMs, there is a need for more research and attention on testing such models.

Traditional software testing techniques face several challenges due to the non-deterministic nature, complex input structures, and the lack of transparency in LLMs' decision-making processes. While techniques like fuzz testing and metamorphic testing may be adapted and extended to these scenarios, they still do not fully address the core complexities of LLMs. These methods may help detect specific issues but often fail to tackle the underlying dynamics of model behavior, such as how models handle ambiguous or unseen inputs. Therefore, there is room for more novel approaches and ideas in testing LLMs, potentially requiring entirely new paradigms that go beyond traditional methods to better evaluate their performance and reliability in diverse, real-world tasks.

\textbf{Testing LLM-as-agent systems.} 
Whether for classification tasks or generation task, LLMs remain far from mature. Yet, LLM-powered applications have already proliferated, with the LLM-as-agent paradigm \cite{xi2023rise,liu2023agentBench} being a prominent example. 
In this scenario, LLMs handle tasks such as cognitive understanding and decision-making to support specific applications. 
This paradigm extends beyond the LLMs themselves, incorporating external components like environments, memory modules for storing interaction history, external knowledge bases for retrieving up-to-date information, and tools or services for task execution. 
Testing the performance of LLM-as-agent systems requires evaluating both traditional software metrics (e.g., response speed, fault tolerance, and reliability) and AI-specific aspects (e.g., compliance, robustness, generalization, trustworthiness, and fairness).

Apart from that, with recent advancements, such as Claude 3.5's computer use and AutoGLM, LLM agents are demonstrating greater autonomy, more sophisticated tool usage, and emerging visual capabilities, evolving toward LLM-centered operating systems. 
Testing in this context involves verifying the model’s ability to effectively utilize external resources, adapt to dynamic environments, and maintain coherence across complex tasks.
Additionally, testing strategies must account for challenging scenarios, including handling unexpected inputs, adapting to novel situations, and recovering gracefully from errors or disruptions. By adopting a holistic testing approach, developers can ensure the functionality and reliability of LLM-as-agent systems in diverse real-world scenarios.


\textbf{Quality assurance of LLM Store (e.g., GPT Store).}
The first two items in this subsection primarily discuss testing from the perspectives of LLMs and LLM agents. 
Here, the discussion will be expanded from the perspective of the LLM ecosystem.
GPTs, as a new form of service based on LLMs, will make the GPT Store a new channel for people to access applications, similar to the Google Play Store or the Apple App Store in the era of mobile applications. 
They are custom versions of GPT tailored for specific purposes, allowing users to create personalized iterations of GPT to better suit their needs \cite{GPTs}. 
These customized GPTs can be designed for various tasks, such as teaching children math, providing assistance in board games, or generating stickers. 
In the context of the mobile app market, tasks related to ensuring the quality and reliability of apps, such as malware detection, privacy violation  detection, and sensitive data leaks detection, remain critical for GPTs. 
However, due to the differences between LLMs and mobile apps, new testing techniques are urgently needed to address these challenges.
Moreover, discrepancies between the descriptions of GPT capabilities and their actual performance may arise, necessitating robust testing procedures to ensure consistency and accuracy. 
Additionally, leveraging user-contributed feedback and reviews on the GPT Store platform could serve as a valuable resource for identifying defects and improving GPT performance, which can take inspiration from previous researches on mining mobile app reviews \cite{gao2018online,dkabrowski2022analysing}.

\subsection{Testing and Quality Assurance for Auto-generated Code}
With the advancement of LLMs and their remarkable performance in code generation tasks, developers are increasingly relying on these models for various coding and debugging tasks. 
Recently, there have been reports of groundbreaking developments, such as Microsoft's creation of an AI programmer named AutoDev \cite{AutoDev}, capable of mastering full-stack skills, which can not only write code and debug, but also train models and even bid for projects on the largest freelancing platform, Upwork.

Automatic code generation is rapidly gaining traction, yet testing practices in this area lag behind. Currently, most approaches utilize code-focused LLMs to assist engineers in writing code, functioning similarly to pair programming in agile methodologies. In this collaborative dynamic, LLMs generate code while engineers review it, relying primarily on traditional testing methods. However, this mismatch between human-driven workflows and AI-driven automation hinders efficiency. Furthermore, challenges such as LLM hallucinations, data poisoning, and other vulnerabilities introduce significant and often hidden risks to code quality.

\textbf{Functional concerns of auto-generated code.} 
The integration of AI-generated code into open-source projects raises concerns regarding code reliability. 
While AI models demonstrate impressive capabilities in code generation, there remain uncertainties about the robustness and correctness of the generated code \cite{Ouyang2023LLM}. 
Liu et al. explores and evaluates the hallucinations in LLM-powered code generation, and categories them into intent conflicting (e.g., local semantic conflicting), context deviation (e.g., generate repetitive statements), and knowledge conflicting (e.g., using un-imported library) \cite{liu2024exploring}.
It is challenging for LLMs to detect and correct hallucinations through prompting. Therefore, it is crucial to develop specialized techniques for detecting and mitigating these issues.
There have been some attempts targeted at specific tasks \cite{luo2024hallucination}, but they have yet to achieve satisfactory results.

\textbf{Non-functional concerns of auto-generated code.} In addition to functional correctness, the non-functional requirements of automatically generated code are also critically important. For example, Zhang et al. proposed EffiBench, a benchmark for assessing the efficiency of automatically generated code \cite{huang_EffiBench_benchmarkingEfficiency}. 
Similarly, maintainability and security concerns, as highlighted by Asare et al. \cite{asare2023github}, are essential aspects of software quality. Developing robust tools and benchmarks to address these non-functional requirements is crucial for ensuring the reliability and practicality of AI-generated code in real-world applications.

In addition to the inherent weaknesses of LLMs that may result in insecure code, external attackers can also introduce security risks into automatically generated code. For example, model publishers might embed malicious backdoors, causing the model to perform normally on standard inputs but generate harmful outputs (i.e., insecure code) when exposed to specific triggered inputs \cite{yang2024stealthy}. These malicious code could lead to data theft, unauthorized software behavior, or other security breaches. Therefore, testing for automatically generated code must also account for such scenarios to ensure robustness against these threats.

\textbf{Testing in terms of system-level auto-generated code.}
The lack of human oversight in the code generation process may lead to the introduction of low-quality code that could compromise the integrity of software systems. 
The sheer volume of AI-generated contributions necessitates scalable testing processes to ensure compliance with  coding standards, adheres to best practices, and complies with project-specific requirements.
To meet these needs, it is crucial to develop testing frameworks specifically tailored to address the nuances of auto-generated code. 
These frameworks should integrate static analysis tools, automated test generation, and dynamic testing techniques to identify and resolve potential issues effectively. 
Moreover, fostering collaboration between AI developers, software engineers, and open-source maintainers is essential to facilitate the seamless integration of AI-generated contributions while upholding the quality and reliability of open-source software projects.

\subsection{\textbf{Collaborative Testing Between Human and AI}}

\textbf{Re-define communication and collaboration between humans and AI systems.}
As described in Section 3, previous studies on human and AI collaboration mainly focused on using AI-related technologies to provide intelligent services for humans, such as recommending suitable issue reports. 
However, with the advancement of technologies like LLMs, AI's capabilities have become more prominent, leading to the emergence of many AI-powered techniques and even AI-powered testers. 
Therefore, we need to re-define communication and collaboration between humans and AI systems during the software testing and quality assurance process.
A widely accepted notion is the human-in-the-loop methodology.

In fact, there are already relevant implementations in the field of software testing. 
For instance, Zamprogno et al. applied it to test assertion generation, where the developer selects the variables they want assertions for, the tool generates assertions only for these variables, and the developer evaluates the relatively small number of generated assertions, ensuring that only useful assertions are persisted in their test cases \cite{zamprogno2022dynamic}. 
Similarly, Geethal et al. applied it to program repair, using it in conjunction with active learning techniques to present test cases with a higher probability of being labeled as failing to the human \cite{geethal2023human}. 
These methods have achieved better performance compared to purely automated techniques.

Meanwhile, the launch of LangGraph, a library to help developers build multi-actor, multi-step, stateful LLM applications, has already supported two ``human-in-the-loop'' features in OpenGPTs: Interrupt and Authorize \cite{human-in-the-loop}. 
The first mode, Interrupt, is the simplest form of control—users monitor the streaming output of the application and manually interrupt it when they deem necessary. 
The second control mode is Authorize, where users pre-define that the application should hand off control to them whenever a particular actor is about to be called. 
This underscores the significance of human involvement, and these two  modes can inspire the creation of other interaction patterns between humans and AI systems for software testing applications.

\section{Conclusion}

The open-collaborative software development paradigm, empowered by internet technology, has significantly transformed the landscape of software testing. 
This paper explores the interconnected dimensions of process, personnel, and technology in the context of modern software testing. 
It also examines the challenges and opportunities presented by the emergence of LLMs. 

In fact, since the release of ChatGPT on November 30, 2022, the AI field has undergone profound changes. What seemed like forward-thinking ideas just a month ago may now appear commonplace due to the rapid emergence of new technologies. In this era of accelerated productivity, it is difficult to predict what will happen in the next five to ten years. Returning to the field of software engineering and software testing, the development of AI is rapidly transforming related research tasks, and we have witnessed how some research areas that were highly relevant just three years ago have now become less frequently discussed. This roadmap is an attempt to reflect on and analyze the past from the vantage point of the present, offering insights into the challenges and opportunities ahead based on our experience. It is clear that software testing will make significant strides in the future, but no one can predict with certainty what those advancements will look like, which is precisely what makes this era so exciting.

\begin{acks}
    This work was supported by the National Natural Science Foundation of China Grant No.62232016, No.62072442, 62402484, 62402483, Youth Innovation Promotion Association CAS, Basic Research Program of ISCAS Grant No. ISCAS-JCZD-202304, Major Program of ISCAS Grant No. ISCAS-ZD-202302 and ISCAS-ZD-202401, Innovation Team 2024 ISCAS (No. 2024-66).
\end{acks}

\normalem
\bibliographystyle{ACM-Reference-Format}
\bibliography{reference,reference-challenge}


\begin{thebibliography}{75}


\ifx \showCODEN    \undefined \def \showCODEN     #1{\unskip}     \fi
\ifx \showDOI      \undefined \def \showDOI       #1{#1}\fi
\ifx \showISBNx    \undefined \def \showISBNx     #1{\unskip}     \fi
\ifx \showISBNxiii \undefined \def \showISBNxiii  #1{\unskip}     \fi
\ifx \showISSN     \undefined \def \showISSN      #1{\unskip}     \fi
\ifx \showLCCN     \undefined \def \showLCCN      #1{\unskip}     \fi
\ifx \shownote     \undefined \def \shownote      #1{#1}          \fi
\ifx \showarticletitle \undefined \def \showarticletitle #1{#1}   \fi
\ifx \showURL      \undefined \def \showURL       {\relax}        \fi
\providecommand\bibfield[2]{#2}
\providecommand\bibinfo[2]{#2}
\providecommand\natexlab[1]{#1}
\providecommand\showeprint[2][]{arXiv:#2}

\bibitem[Alagarsamy et~al\mbox{.}(2023)]%
        {23a3testTestGeneration}
\bibfield{author}{\bibinfo{person}{Saranya Alagarsamy}, \bibinfo{person}{Chakkrit Tantithamthavorn}, {and} \bibinfo{person}{Aldeida Aleti}.} \bibinfo{year}{2023}\natexlab{}.
\newblock \showarticletitle{A3Test: Assertion-Augmented Automated Test Case Generation}.
\newblock \bibinfo{journal}{\emph{arXiv preprint arXiv:2302.10352}} (\bibinfo{year}{2023}).
\newblock


\bibitem[Anvik et~al\mbox{.}(2006)]%
        {anvik2006should}
\bibfield{author}{\bibinfo{person}{John Anvik}, \bibinfo{person}{Lyndon Hiew}, {and} \bibinfo{person}{Gail~C Murphy}.} \bibinfo{year}{2006}\natexlab{}.
\newblock \showarticletitle{Who should fix this bug?}. In \bibinfo{booktitle}{\emph{Proceedings of the 28th international conference on Software engineering}}. \bibinfo{pages}{361--370}.
\newblock


\bibitem[Asare et~al\mbox{.}(2023)]%
        {asare2023github}
\bibfield{author}{\bibinfo{person}{Owura Asare}, \bibinfo{person}{Meiyappan Nagappan}, {and} \bibinfo{person}{N Asokan}.} \bibinfo{year}{2023}\natexlab{}.
\newblock \showarticletitle{Is github’s copilot as bad as humans at introducing vulnerabilities in code?}
\newblock \bibinfo{journal}{\emph{Empirical Software Engineering}} \bibinfo{volume}{28}, \bibinfo{number}{6} (\bibinfo{year}{2023}), \bibinfo{pages}{129}.
\newblock


\bibitem[Bagherzadeh et~al\mbox{.}(2022)]%
        {Bagherzadeh2022Reinforcement}
\bibfield{author}{\bibinfo{person}{Mojtaba Bagherzadeh}, \bibinfo{person}{Nafiseh Kahani}, {and} \bibinfo{person}{Lionel~C. Briand}.} \bibinfo{year}{2022}\natexlab{}.
\newblock \showarticletitle{Reinforcement Learning for Test Case Prioritization}.
\newblock \bibinfo{journal}{\emph{{IEEE} Trans. Software Eng.}} \bibinfo{volume}{48}, \bibinfo{number}{8} (\bibinfo{year}{2022}), \bibinfo{pages}{2836--2856}.
\newblock
\urldef\tempurl%
\url{https://doi.org/10.1109/TSE.2021.3070549}
\showDOI{\tempurl}


\bibitem[Bertolino et~al\mbox{.}(2020)]%
        {Bertolino2020Learning}
\bibfield{author}{\bibinfo{person}{Antonia Bertolino}, \bibinfo{person}{Antonio Guerriero}, \bibinfo{person}{Breno Miranda}, \bibinfo{person}{Roberto Pietrantuono}, {and} \bibinfo{person}{Stefano Russo}.} \bibinfo{year}{2020}\natexlab{}.
\newblock \showarticletitle{Learning-to-rank vs ranking-to-learn: strategies for regression testing in continuous integration}. In \bibinfo{booktitle}{\emph{{ICSE} '20: 42nd International Conference on Software Engineering, Seoul, South Korea, 27 June - 19 July, 2020}}, \bibfield{editor}{\bibinfo{person}{Gregg Rothermel} {and} \bibinfo{person}{Doo{-}Hwan Bae}} (Eds.). \bibinfo{publisher}{{ACM}}, \bibinfo{pages}{1--12}.
\newblock
\urldef\tempurl%
\url{https://doi.org/10.1145/3377811.3380369}
\showDOI{\tempurl}


\bibitem[Booch and Brown(2003)]%
        {booch2003collaborative}
\bibfield{author}{\bibinfo{person}{Grady Booch} {and} \bibinfo{person}{Alan~W Brown}.} \bibinfo{year}{2003}\natexlab{}.
\newblock \showarticletitle{Collaborative development environments}.
\newblock \bibinfo{journal}{\emph{Adv. Comput.}} \bibinfo{volume}{59}, \bibinfo{number}{1} (\bibinfo{year}{2003}), \bibinfo{pages}{1--27}.
\newblock


\bibitem[Busjaeger and Xie(2016)]%
        {Busjaeger2016Learning}
\bibfield{author}{\bibinfo{person}{Benjamin Busjaeger} {and} \bibinfo{person}{Tao Xie}.} \bibinfo{year}{2016}\natexlab{}.
\newblock \showarticletitle{Learning for test prioritization: an industrial case study}. In \bibinfo{booktitle}{\emph{Proceedings of the 24th {ACM} {SIGSOFT} International Symposium on Foundations of Software Engineering, {FSE} 2016, Seattle, WA, USA, November 13-18, 2016}}, \bibfield{editor}{\bibinfo{person}{Thomas Zimmermann}, \bibinfo{person}{Jane Cleland{-}Huang}, {and} \bibinfo{person}{Zhendong Su}} (Eds.). \bibinfo{publisher}{{ACM}}, \bibinfo{pages}{975--980}.
\newblock
\urldef\tempurl%
\url{https://doi.org/10.1145/2950290.2983954}
\showDOI{\tempurl}


\bibitem[Chen et~al\mbox{.}(2024)]%
        {chen2024mllmasajudgeassessingmultimodalllmasajudge}
\bibfield{author}{\bibinfo{person}{Dongping Chen}, \bibinfo{person}{Ruoxi Chen}, \bibinfo{person}{Shilin Zhang}, \bibinfo{person}{Yinuo Liu}, \bibinfo{person}{Yaochen Wang}, \bibinfo{person}{Huichi Zhou}, \bibinfo{person}{Qihui Zhang}, \bibinfo{person}{Yao Wan}, \bibinfo{person}{Pan Zhou}, {and} \bibinfo{person}{Lichao Sun}.} \bibinfo{year}{2024}\natexlab{}.
\newblock \bibinfo{title}{MLLM-as-a-Judge: Assessing Multimodal LLM-as-a-Judge with Vision-Language Benchmark}.
\newblock
\newblock
\showeprint[arxiv, accepted to icml 2024]{2402.04788}~[cs.CL]
\urldef\tempurl%
\url{https://arxiv.org/abs/2402.04788}
\showURL{%
\tempurl}


\bibitem[Chen et~al\mbox{.}(2020)]%
        {chen2020improving}
\bibfield{author}{\bibinfo{person}{Yan Chen}, \bibinfo{person}{Maulishree Pandey}, \bibinfo{person}{Jean~Y Song}, \bibinfo{person}{Walter~S Lasecki}, {and} \bibinfo{person}{Steve Oney}.} \bibinfo{year}{2020}\natexlab{}.
\newblock \showarticletitle{Improving crowd-supported gui testing with structural guidance}. In \bibinfo{booktitle}{\emph{Proceedings of the 2020 CHI Conference on Human Factors in Computing Systems}}. \bibinfo{pages}{1--13}.
\newblock


\bibitem[Crowston et~al\mbox{.}(2008)]%
        {crowston2008free}
\bibfield{author}{\bibinfo{person}{Kevin Crowston}, \bibinfo{person}{Kangning Wei}, \bibinfo{person}{James Howison}, {and} \bibinfo{person}{Andrea Wiggins}.} \bibinfo{year}{2008}\natexlab{}.
\newblock \showarticletitle{Free/Libre open-source software development: What we know and what we do not know}.
\newblock \bibinfo{journal}{\emph{ACM Computing Surveys (CSUR)}} \bibinfo{volume}{44}, \bibinfo{number}{2} (\bibinfo{year}{2008}), \bibinfo{pages}{1--35}.
\newblock


\bibitem[D{\k{a}}browski et~al\mbox{.}(2022)]%
        {dkabrowski2022analysing}
\bibfield{author}{\bibinfo{person}{Jacek D{\k{a}}browski}, \bibinfo{person}{Emmanuel Letier}, \bibinfo{person}{Anna Perini}, {and} \bibinfo{person}{Angelo Susi}.} \bibinfo{year}{2022}\natexlab{}.
\newblock \showarticletitle{Analysing app reviews for software engineering: a systematic literature review}.
\newblock \bibinfo{journal}{\emph{Empirical Software Engineering}} \bibinfo{volume}{27}, \bibinfo{number}{2} (\bibinfo{year}{2022}), \bibinfo{pages}{43}.
\newblock


\bibitem[Deng et~al\mbox{.}(2023a)]%
        {19fuzzDeepLearningLibraries}
\bibfield{author}{\bibinfo{person}{Yinlin Deng}, \bibinfo{person}{Chunqiu~Steven Xia}, \bibinfo{person}{Chenyuan Yang}, \bibinfo{person}{Shizhuo~Dylan Zhang}, \bibinfo{person}{Shujing Yang}, {and} \bibinfo{person}{Lingming Zhang}.} \bibinfo{year}{2023}\natexlab{a}.
\newblock \showarticletitle{Large language models are edge-case fuzzers: Testing deep learning libraries via fuzzgpt}.
\newblock \bibinfo{journal}{\emph{ICSE 2024}} (\bibinfo{year}{2023}).
\newblock


\bibitem[Deng et~al\mbox{.}(2023b)]%
        {13fuzzDeepLearningLibraries}
\bibfield{author}{\bibinfo{person}{Yinlin Deng}, \bibinfo{person}{Chunqiu~Steven Xia}, \bibinfo{person}{Chenyuan Yang}, \bibinfo{person}{Shizhuo~Dylan Zhang}, \bibinfo{person}{Shujing Yang}, {and} \bibinfo{person}{Lingming Zhang}.} \bibinfo{year}{2023}\natexlab{b}.
\newblock \showarticletitle{Large Language Models are Zero Shot Fuzzers: Fuzzing Deep Learning Libraries via Large Language Models}.
\newblock \bibinfo{journal}{\emph{ISSTA 2023}} (\bibinfo{year}{2023}).
\newblock


\bibitem[Dipongkor and Moran(2023)]%
        {Dipongkor3023comparative}
\bibfield{author}{\bibinfo{person}{Atish~Kumar Dipongkor} {and} \bibinfo{person}{Kevin Moran}.} \bibinfo{year}{2023}\natexlab{}.
\newblock \showarticletitle{A Comparative Study of Transformer-Based Neural Text Representation Techniques on Bug Triaging}. In \bibinfo{booktitle}{\emph{38th {IEEE/ACM} International Conference on Automated Software Engineering, {ASE} 2023, Luxembourg, September 11-15, 2023}}. \bibinfo{publisher}{{IEEE}}, \bibinfo{pages}{1012--1023}.
\newblock
\urldef\tempurl%
\url{https://doi.org/10.1109/ASE56229.2023.00217}
\showDOI{\tempurl}


\bibitem[Dwarakanath et~al\mbox{.}(2018)]%
        {DwarakanathASRB2018Identifying}
\bibfield{author}{\bibinfo{person}{Anurag Dwarakanath}, \bibinfo{person}{Manish Ahuja}, \bibinfo{person}{Samarth Sikand}, \bibinfo{person}{Raghotham~M. Rao}, \bibinfo{person}{R.~P. Jagadeesh~Chandra Bose}, \bibinfo{person}{Neville Dubash}, {and} \bibinfo{person}{Sanjay Podder}.} \bibinfo{year}{2018}\natexlab{}.
\newblock \showarticletitle{Identifying implementation bugs in machine learning based image classifiers using metamorphic testing}. In \bibinfo{booktitle}{\emph{Proceedings of the 27th {ACM} {SIGSOFT} International Symposium on Software Testing and Analysis, {ISSTA} 2018, Amsterdam, The Netherlands, July 16-21, 2018}}, \bibfield{editor}{\bibinfo{person}{Frank Tip} {and} \bibinfo{person}{Eric Bodden}} (Eds.). \bibinfo{publisher}{{ACM}}, \bibinfo{pages}{118--128}.
\newblock
\urldef\tempurl%
\url{https://doi.org/10.1145/3213846.3213858}
\showDOI{\tempurl}


\bibitem[Elbaum et~al\mbox{.}(2014)]%
        {Elbaum2014techniques}
\bibfield{author}{\bibinfo{person}{Sebastian~G. Elbaum}, \bibinfo{person}{Gregg Rothermel}, {and} \bibinfo{person}{John Penix}.} \bibinfo{year}{2014}\natexlab{}.
\newblock \showarticletitle{Techniques for improving regression testing in continuous integration development environments}. In \bibinfo{booktitle}{\emph{Proceedings of the 22nd {ACM} {SIGSOFT} International Symposium on Foundations of Software Engineering, (FSE-22), Hong Kong, China, November 16 - 22, 2014}}, \bibfield{editor}{\bibinfo{person}{Shing{-}Chi Cheung}, \bibinfo{person}{Alessandro Orso}, {and} \bibinfo{person}{Margaret{-}Anne~D. Storey}} (Eds.). \bibinfo{publisher}{{ACM}}, \bibinfo{pages}{235--245}.
\newblock
\urldef\tempurl%
\url{https://doi.org/10.1145/2635868.2635910}
\showDOI{\tempurl}


\bibitem[Gao et~al\mbox{.}(2018)]%
        {gao2018online}
\bibfield{author}{\bibinfo{person}{Cuiyun Gao}, \bibinfo{person}{Jichuan Zeng}, \bibinfo{person}{Michael~R Lyu}, {and} \bibinfo{person}{Irwin King}.} \bibinfo{year}{2018}\natexlab{}.
\newblock \showarticletitle{Online app review analysis for identifying emerging issues}. In \bibinfo{booktitle}{\emph{Proceedings of the 40th International Conference on Software Engineering}}. \bibinfo{pages}{48--58}.
\newblock


\bibitem[Geethal et~al\mbox{.}(2023)]%
        {geethal2023human}
\bibfield{author}{\bibinfo{person}{Charaka Geethal}, \bibinfo{person}{Marcel B{\"o}hme}, {and} \bibinfo{person}{Van-Thuan Pham}.} \bibinfo{year}{2023}\natexlab{}.
\newblock \showarticletitle{Human-In-The-Loop Automatic Program Repair}.
\newblock \bibinfo{journal}{\emph{IEEE Transactions on Software Engineering}} (\bibinfo{year}{2023}).
\newblock


\bibitem[Gruzd and Haythornthwaite(2013)]%
        {gruzd2013enabling}
\bibfield{author}{\bibinfo{person}{Anatoliy Gruzd} {and} \bibinfo{person}{Caroline Haythornthwaite}.} \bibinfo{year}{2013}\natexlab{}.
\newblock \showarticletitle{Enabling community through social media}.
\newblock \bibinfo{journal}{\emph{Journal of medical Internet research}} \bibinfo{volume}{15}, \bibinfo{number}{10} (\bibinfo{year}{2013}), \bibinfo{pages}{e248}.
\newblock


\bibitem[Haghighatkhah et~al\mbox{.}(2018)]%
        {Haghighatkhah2018Test}
\bibfield{author}{\bibinfo{person}{Alireza Haghighatkhah}, \bibinfo{person}{Mika M{\"{a}}ntyl{\"{a}}}, \bibinfo{person}{Markku Oivo}, {and} \bibinfo{person}{Pasi Kuvaja}.} \bibinfo{year}{2018}\natexlab{}.
\newblock \showarticletitle{Test prioritization in continuous integration environments}.
\newblock \bibinfo{journal}{\emph{J. Syst. Softw.}}  \bibinfo{volume}{146} (\bibinfo{year}{2018}), \bibinfo{pages}{80--98}.
\newblock
\urldef\tempurl%
\url{https://doi.org/10.1016/J.JSS.2018.08.061}
\showDOI{\tempurl}


\bibitem[Hassan(2008)]%
        {hassan2008road}
\bibfield{author}{\bibinfo{person}{Ahmed~E Hassan}.} \bibinfo{year}{2008}\natexlab{}.
\newblock \showarticletitle{The road ahead for mining software repositories}. In \bibinfo{booktitle}{\emph{2008 frontiers of software maintenance}}. IEEE, \bibinfo{pages}{48--57}.
\newblock


\bibitem[Huang et~al\mbox{.}(2024)]%
        {huang_EffiBench_benchmarkingEfficiency}
\bibfield{author}{\bibinfo{person}{Dong Huang}, \bibinfo{person}{Jie~M. Zhang}, \bibinfo{person}{Yuhao Qing}, {and} \bibinfo{person}{Heming Cui}.} \bibinfo{year}{2024}\natexlab{}.
\newblock \showarticletitle{EffiBench: Benchmarking the Efficiency of Automatically Generated Code}.
\newblock \bibinfo{journal}{\emph{CoRR}}  \bibinfo{volume}{abs/2402.02037} (\bibinfo{year}{2024}).
\newblock
\urldef\tempurl%
\url{https://doi.org/10.48550/ARXIV.2402.02037}
\showDOI{\tempurl}
\showeprint[arXiv]{2402.02037}


\bibitem[Huang et~al\mbox{.}(2020)]%
        {huang2020questfor}
\bibfield{author}{\bibinfo{person}{Yuekai Huang}, \bibinfo{person}{Junjie Wang}, \bibinfo{person}{Song Wang}, \bibinfo{person}{Zhe Liu}, \bibinfo{person}{Yuanzhe Hu}, {and} \bibinfo{person}{Qing Wang}.} \bibinfo{year}{2020}\natexlab{}.
\newblock \showarticletitle{Quest for the Golden Approach: An Experimental Evaluation of Duplicate Crowdtesting Reports Detection}. In \bibinfo{booktitle}{\emph{{ESEM} '20: {ACM} / {IEEE} International Symposium on Empirical Software Engineering and Measurement, Bari, Italy, October 5-7, 2020}}, \bibfield{editor}{\bibinfo{person}{Maria~Teresa Baldassarre}, \bibinfo{person}{Filippo Lanubile}, \bibinfo{person}{Marcos Kalinowski}, {and} \bibinfo{person}{Federica Sarro}} (Eds.). \bibinfo{publisher}{{ACM}}, \bibinfo{pages}{17:1--17:12}.
\newblock
\urldef\tempurl%
\url{https://doi.org/10.1145/3382494.3410694}
\showDOI{\tempurl}


\bibitem[Jeong et~al\mbox{.}(2009)]%
        {jeong2009improving}
\bibfield{author}{\bibinfo{person}{Gaeul Jeong}, \bibinfo{person}{Sunghun Kim}, {and} \bibinfo{person}{Thomas Zimmermann}.} \bibinfo{year}{2009}\natexlab{}.
\newblock \showarticletitle{Improving bug triage with bug tossing graphs}. In \bibinfo{booktitle}{\emph{Proceedings of the 7th joint meeting of the European software engineering conference and the ACM SIGSOFT symposium on The foundations of software engineering}}. \bibinfo{pages}{111--120}.
\newblock


\bibitem[Jiang et~al\mbox{.}(2024)]%
        {jiang2024towards}
\bibfield{author}{\bibinfo{person}{Zongze Jiang}, \bibinfo{person}{Ming Wen}, \bibinfo{person}{Jialun Cao}, \bibinfo{person}{Xuanhua Shi}, {and} \bibinfo{person}{Hai Jin}.} \bibinfo{year}{2024}\natexlab{}.
\newblock \showarticletitle{Towards Understanding the Effectiveness of Large Language Models on Directed Test Input Generation}. In \bibinfo{booktitle}{\emph{Proceedings of the 39th IEEE/ACM International Conference on Automated Software Engineering}}. \bibinfo{pages}{1408--1420}.
\newblock


\bibitem[langchain(2024)]%
        {human-in-the-loop}
\bibfield{author}{\bibinfo{person}{langchain}.} \bibinfo{year}{2024}\natexlab{}.
\newblock
\newblock
\urldef\tempurl%
\url{https://blog.langchain.dev/human-in-the-loop-with-opengpts-and-langgraph/}
\showURL{%
\tempurl}


\bibitem[Lee et~al\mbox{.}(2017)]%
        {lee2017applying}
\bibfield{author}{\bibinfo{person}{Sun{-}Ro Lee}, \bibinfo{person}{Min{-}Jae Heo}, \bibinfo{person}{Chan{-}Gun Lee}, \bibinfo{person}{Milhan Kim}, {and} \bibinfo{person}{Gaeul Jeong}.} \bibinfo{year}{2017}\natexlab{}.
\newblock \showarticletitle{Applying deep learning based automatic bug triager to industrial projects}. In \bibinfo{booktitle}{\emph{Proceedings of the 2017 11th Joint Meeting on Foundations of Software Engineering, {ESEC/FSE} 2017, Paderborn, Germany, September 4-8, 2017}}, \bibfield{editor}{\bibinfo{person}{Eric Bodden}, \bibinfo{person}{Wilhelm Sch{\"{a}}fer}, \bibinfo{person}{Arie van Deursen}, {and} \bibinfo{person}{Andrea Zisman}} (Eds.). \bibinfo{publisher}{{ACM}}, \bibinfo{pages}{926--931}.
\newblock
\urldef\tempurl%
\url{https://doi.org/10.1145/3106237.3117776}
\showDOI{\tempurl}


\bibitem[Lima and Vergilio(2022)]%
        {Lima2022Multi-Armed}
\bibfield{author}{\bibinfo{person}{Jackson A.~Prado Lima} {and} \bibinfo{person}{Silvia~Regina Vergilio}.} \bibinfo{year}{2022}\natexlab{}.
\newblock \showarticletitle{A Multi-Armed Bandit Approach for Test Case Prioritization in Continuous Integration Environments}.
\newblock \bibinfo{journal}{\emph{{IEEE} Trans. Software Eng.}} \bibinfo{volume}{48}, \bibinfo{number}{2} (\bibinfo{year}{2022}), \bibinfo{pages}{453--465}.
\newblock
\urldef\tempurl%
\url{https://doi.org/10.1109/TSE.2020.2992428}
\showDOI{\tempurl}


\bibitem[Liu et~al\mbox{.}(2024c)]%
        {liu2024exploring}
\bibfield{author}{\bibinfo{person}{Fang Liu}, \bibinfo{person}{Yang Liu}, \bibinfo{person}{Lin Shi}, \bibinfo{person}{Houkun Huang}, \bibinfo{person}{Ruifeng Wang}, \bibinfo{person}{Zhen Yang}, {and} \bibinfo{person}{Li Zhang}.} \bibinfo{year}{2024}\natexlab{c}.
\newblock \showarticletitle{Exploring and evaluating hallucinations in llm-powered code generation}.
\newblock \bibinfo{journal}{\emph{arXiv preprint arXiv:2404.00971}} (\bibinfo{year}{2024}).
\newblock


\bibitem[Liu et~al\mbox{.}(2023b)]%
        {liu2023agentBench}
\bibfield{author}{\bibinfo{person}{Xiao Liu}, \bibinfo{person}{Hao Yu}, \bibinfo{person}{Hanchen Zhang}, \bibinfo{person}{Yifan Xu}, \bibinfo{person}{Xuanyu Lei}, \bibinfo{person}{Hanyu Lai}, \bibinfo{person}{Yu Gu}, \bibinfo{person}{Hangliang Ding}, \bibinfo{person}{Kaiwen Men}, \bibinfo{person}{Kejuan Yang}, \bibinfo{person}{Shudan Zhang}, \bibinfo{person}{Xiang Deng}, \bibinfo{person}{Aohan Zeng}, \bibinfo{person}{Zhengxiao Du}, \bibinfo{person}{Chenhui Zhang}, \bibinfo{person}{Sheng Shen}, \bibinfo{person}{Tianjun Zhang}, \bibinfo{person}{Yu Su}, \bibinfo{person}{Huan Sun}, \bibinfo{person}{Minlie Huang}, \bibinfo{person}{Yuxiao Dong}, {and} \bibinfo{person}{Jie Tang}.} \bibinfo{year}{2023}\natexlab{b}.
\newblock \showarticletitle{AgentBench: Evaluating LLMs as Agents}.
\newblock \bibinfo{journal}{\emph{CoRR}}  \bibinfo{volume}{abs/2308.03688} (\bibinfo{year}{2023}).
\newblock
\urldef\tempurl%
\url{https://doi.org/10.48550/ARXIV.2308.03688}
\showDOI{\tempurl}
\showeprint[arXiv]{2308.03688}


\bibitem[Liu et~al\mbox{.}(2023a)]%
        {26fillBlank}
\bibfield{author}{\bibinfo{person}{Zhe Liu}, \bibinfo{person}{Chunyang Chen}, \bibinfo{person}{Junjie Wang}, \bibinfo{person}{Xing Che}, \bibinfo{person}{Yuekai Huang}, \bibinfo{person}{Jun Hu}, {and} \bibinfo{person}{Qing Wang}.} \bibinfo{year}{2023}\natexlab{a}.
\newblock \showarticletitle{Fill in the Blank: Context-aware Automated Text Input Generation for Mobile GUI Testing}.
\newblock \bibinfo{journal}{\emph{ICSE 2023}} (\bibinfo{year}{2023}).
\newblock


\bibitem[Liu et~al\mbox{.}(2024a)]%
        {60GUITesting}
\bibfield{author}{\bibinfo{person}{Zhe Liu}, \bibinfo{person}{Chunyang Chen}, \bibinfo{person}{Junjie Wang}, \bibinfo{person}{Mengzhuo Chen}, \bibinfo{person}{Boyu Wu}, \bibinfo{person}{Xing Che}, \bibinfo{person}{Dandan Wang}, {and} \bibinfo{person}{Qing Wang}.} \bibinfo{year}{2024}\natexlab{a}.
\newblock \showarticletitle{Make {LLM} a Testing Expert: Bringing Human-like Interaction to Mobile {GUI} Testing via Functionality-aware Decisions}.
\newblock \bibinfo{journal}{\emph{ICSE 2024}} (\bibinfo{year}{2024}).
\newblock


\bibitem[Liu et~al\mbox{.}(2021)]%
        {liu2021owleyes}
\bibfield{author}{\bibinfo{person}{Zhe Liu}, \bibinfo{person}{Chunyang Chen}, \bibinfo{person}{Junjie Wang}, \bibinfo{person}{Yuekai Huang}, \bibinfo{person}{Jun Hu}, {and} \bibinfo{person}{Qing Wang}.} \bibinfo{year}{2021}\natexlab{}.
\newblock \showarticletitle{Owl eyes: spotting UI display issues via visual understanding}. In \bibinfo{booktitle}{\emph{Proceedings of the 35th IEEE/ACM International Conference on Automated Software Engineering}} (Virtual Event, Australia) \emph{(\bibinfo{series}{ASE '20})}. \bibinfo{publisher}{Association for Computing Machinery}, \bibinfo{address}{New York, NY, USA}, \bibinfo{pages}{398–409}.
\newblock
\showISBNx{9781450367684}
\urldef\tempurl%
\url{https://doi.org/10.1145/3324884.3416547}
\showDOI{\tempurl}


\bibitem[Liu et~al\mbox{.}(2022)]%
        {liu2022guided}
\bibfield{author}{\bibinfo{person}{Zhe Liu}, \bibinfo{person}{Chunyang Chen}, \bibinfo{person}{Junjie Wang}, \bibinfo{person}{Yuekai Huang}, \bibinfo{person}{Jun Hu}, {and} \bibinfo{person}{Qing Wang}.} \bibinfo{year}{2022}\natexlab{}.
\newblock \showarticletitle{Guided bug crush: Assist manual gui testing of android apps via hint moves}. In \bibinfo{booktitle}{\emph{Proceedings of the 2022 CHI Conference on Human Factors in Computing Systems}}. \bibinfo{pages}{1--14}.
\newblock


\bibitem[Liu et~al\mbox{.}(2024b)]%
        {liu2024visiondrivenautomatedmobilegui}
\bibfield{author}{\bibinfo{person}{Zhe Liu}, \bibinfo{person}{Cheng Li}, \bibinfo{person}{Chunyang Chen}, \bibinfo{person}{Junjie Wang}, \bibinfo{person}{Boyu Wu}, \bibinfo{person}{Yawen Wang}, \bibinfo{person}{Jun Hu}, {and} \bibinfo{person}{Qing Wang}.} \bibinfo{year}{2024}\natexlab{b}.
\newblock \bibinfo{title}{Vision-driven Automated Mobile GUI Testing via Multimodal Large Language Model}.
\newblock
\newblock
\showeprint[arxiv]{2407.03037}~[cs.SE]
\urldef\tempurl%
\url{https://arxiv.org/abs/2407.03037}
\showURL{%
\tempurl}


\bibitem[Lozhkov et~al\mbox{.}(2024)]%
        {lozhkov2024starcoder2stackv2}
\bibfield{author}{\bibinfo{person}{Anton Lozhkov}, \bibinfo{person}{Raymond Li}, \bibinfo{person}{Loubna~Ben Allal}, \bibinfo{person}{Federico Cassano}, \bibinfo{person}{Joel Lamy-Poirier}, \bibinfo{person}{Nouamane Tazi}, \bibinfo{person}{Ao Tang}, \bibinfo{person}{Dmytro Pykhtar}, \bibinfo{person}{Jiawei Liu}, \bibinfo{person}{Yuxiang Wei}, \bibinfo{person}{Tianyang Liu}, \bibinfo{person}{Max Tian}, \bibinfo{person}{Denis Kocetkov}, \bibinfo{person}{Arthur Zucker}, \bibinfo{person}{Younes Belkada}, \bibinfo{person}{Zijian Wang}, \bibinfo{person}{Qian Liu}, \bibinfo{person}{Dmitry Abulkhanov}, \bibinfo{person}{Indraneil Paul}, \bibinfo{person}{Zhuang Li}, \bibinfo{person}{Wen-Ding Li}, \bibinfo{person}{Megan Risdal}, \bibinfo{person}{Jia Li}, \bibinfo{person}{Jian Zhu}, \bibinfo{person}{Terry~Yue Zhuo}, \bibinfo{person}{Evgenii Zheltonozhskii}, \bibinfo{person}{Nii Osae~Osae Dade}, \bibinfo{person}{Wenhao Yu}, \bibinfo{person}{Lucas Krauß}, \bibinfo{person}{Naman Jain}, \bibinfo{person}{Yixuan Su},
  \bibinfo{person}{Xuanli He}, \bibinfo{person}{Manan Dey}, \bibinfo{person}{Edoardo Abati}, \bibinfo{person}{Yekun Chai}, \bibinfo{person}{Niklas Muennighoff}, \bibinfo{person}{Xiangru Tang}, \bibinfo{person}{Muhtasham Oblokulov}, \bibinfo{person}{Christopher Akiki}, \bibinfo{person}{Marc Marone}, \bibinfo{person}{Chenghao Mou}, \bibinfo{person}{Mayank Mishra}, \bibinfo{person}{Alex Gu}, \bibinfo{person}{Binyuan Hui}, \bibinfo{person}{Tri Dao}, \bibinfo{person}{Armel Zebaze}, \bibinfo{person}{Olivier Dehaene}, \bibinfo{person}{Nicolas Patry}, \bibinfo{person}{Canwen Xu}, \bibinfo{person}{Julian McAuley}, \bibinfo{person}{Han Hu}, \bibinfo{person}{Torsten Scholak}, \bibinfo{person}{Sebastien Paquet}, \bibinfo{person}{Jennifer Robinson}, \bibinfo{person}{Carolyn~Jane Anderson}, \bibinfo{person}{Nicolas Chapados}, \bibinfo{person}{Mostofa Patwary}, \bibinfo{person}{Nima Tajbakhsh}, \bibinfo{person}{Yacine Jernite}, \bibinfo{person}{Carlos~Muñoz Ferrandis}, \bibinfo{person}{Lingming Zhang},
  \bibinfo{person}{Sean Hughes}, \bibinfo{person}{Thomas Wolf}, \bibinfo{person}{Arjun Guha}, \bibinfo{person}{Leandro von Werra}, {and} \bibinfo{person}{Harm de Vries}.} \bibinfo{year}{2024}\natexlab{}.
\newblock \bibinfo{title}{StarCoder 2 and The Stack v2: The Next Generation}.
\newblock
\newblock
\showeprint[arxiv]{2402.19173}~[cs.SE]
\urldef\tempurl%
\url{https://arxiv.org/abs/2402.19173}
\showURL{%
\tempurl}


\bibitem[Luo et~al\mbox{.}(2024)]%
        {luo2024hallucination}
\bibfield{author}{\bibinfo{person}{Junliang Luo}, \bibinfo{person}{Tianyu Li}, \bibinfo{person}{Di Wu}, \bibinfo{person}{Michael Jenkin}, \bibinfo{person}{Steve Liu}, {and} \bibinfo{person}{Gregory Dudek}.} \bibinfo{year}{2024}\natexlab{}.
\newblock \showarticletitle{Hallucination detection and hallucination mitigation: An investigation}.
\newblock \bibinfo{journal}{\emph{arXiv preprint arXiv:2401.08358}} (\bibinfo{year}{2024}).
\newblock


\bibitem[Luu et~al\mbox{.}(2023)]%
        {luu2023chatgptadvancesoftwaretesting}
\bibfield{author}{\bibinfo{person}{Quang-Hung Luu}, \bibinfo{person}{Huai Liu}, {and} \bibinfo{person}{Tsong~Yueh Chen}.} \bibinfo{year}{2023}\natexlab{}.
\newblock \bibinfo{title}{Can ChatGPT advance software testing intelligence? An experience report on metamorphic testing}.
\newblock
\newblock
\showeprint[arxiv]{2310.19204}~[cs.SE]
\urldef\tempurl%
\url{https://arxiv.org/abs/2310.19204}
\showURL{%
\tempurl}


\bibitem[Mao et~al\mbox{.}(2017)]%
        {mao2017crowdintelligence}
\bibfield{author}{\bibinfo{person}{Ke Mao}, \bibinfo{person}{Mark Harman}, {and} \bibinfo{person}{Yue Jia}.} \bibinfo{year}{2017}\natexlab{}.
\newblock \showarticletitle{Crowd intelligence enhances automated mobile testing}. In \bibinfo{booktitle}{\emph{2017 32nd IEEE/ACM International Conference on Automated Software Engineering (ASE)}}. \bibinfo{pages}{16--26}.
\newblock
\urldef\tempurl%
\url{https://doi.org/10.1109/ASE.2017.8115614}
\showDOI{\tempurl}


\bibitem[Microsoft(2024)]%
        {AutoDev}
\bibfield{author}{\bibinfo{person}{Microsoft}.} \bibinfo{year}{2024}\natexlab{}.
\newblock \bibinfo{title}{AutoDev}.
\newblock
\newblock
\urldef\tempurl%
\url{https://visualstudiomagazine.com/Articles/2024/03/20/autodev.aspx}
\showURL{%
\tempurl}


\bibitem[Nguyen et~al\mbox{.}(2012)]%
        {Nguyen2012duplicate}
\bibfield{author}{\bibinfo{person}{Anh~Tuan Nguyen}, \bibinfo{person}{Tung~Thanh Nguyen}, \bibinfo{person}{Tien~N. Nguyen}, \bibinfo{person}{David Lo}, {and} \bibinfo{person}{Chengnian Sun}.} \bibinfo{year}{2012}\natexlab{}.
\newblock \showarticletitle{Duplicate bug report detection with a combination of information retrieval and topic modeling}. In \bibinfo{booktitle}{\emph{{IEEE/ACM} International Conference on Automated Software Engineering, ASE'12, Essen, Germany, September 3-7, 2012}}, \bibfield{editor}{\bibinfo{person}{Michael Goedicke}, \bibinfo{person}{Tim Menzies}, {and} \bibinfo{person}{Motoshi Saeki}} (Eds.). \bibinfo{publisher}{{ACM}}, \bibinfo{pages}{70--79}.
\newblock
\urldef\tempurl%
\url{https://doi.org/10.1145/2351676.2351687}
\showDOI{\tempurl}


\bibitem[OpenAI(2024)]%
        {GPTs}
\bibfield{author}{\bibinfo{person}{OpenAI}.} \bibinfo{year}{2024}\natexlab{}.
\newblock \bibinfo{title}{GPTs}.
\newblock
\newblock
\urldef\tempurl%
\url{https://openai.com/blog/introducing-gpts}
\showURL{%
\tempurl}


\bibitem[Orso and Rothermel(2014)]%
        {Orso2014softwareTesting}
\bibfield{author}{\bibinfo{person}{Alessandro Orso} {and} \bibinfo{person}{Gregg Rothermel}.} \bibinfo{year}{2014}\natexlab{}.
\newblock \showarticletitle{Software testing: a research travelogue {(2000-2014)}}. In \bibinfo{booktitle}{\emph{Proceedings of the on Future of Software Engineering, {FOSE} 2014, Hyderabad, India, May 31 - June 7, 2014}}, \bibfield{editor}{\bibinfo{person}{James~D. Herbsleb} {and} \bibinfo{person}{Matthew~B. Dwyer}} (Eds.). \bibinfo{publisher}{{ACM}}, \bibinfo{pages}{117--132}.
\newblock
\urldef\tempurl%
\url{https://doi.org/10.1145/2593882.2593885}
\showDOI{\tempurl}


\bibitem[Ouyang et~al\mbox{.}(2023)]%
        {Ouyang2023LLM}
\bibfield{author}{\bibinfo{person}{Shuyin Ouyang}, \bibinfo{person}{Jie~M. Zhang}, \bibinfo{person}{Mark Harman}, {and} \bibinfo{person}{Meng Wang}.} \bibinfo{year}{2023}\natexlab{}.
\newblock \showarticletitle{{LLM} is Like a Box of Chocolates: the Non-determinism of ChatGPT in Code Generation}.
\newblock \bibinfo{journal}{\emph{CoRR}}  \bibinfo{volume}{abs/2308.02828} (\bibinfo{year}{2023}).
\newblock
\urldef\tempurl%
\url{https://doi.org/10.48550/ARXIV.2308.02828}
\showDOI{\tempurl}
\showeprint[arXiv]{2308.02828}


\bibitem[Pei et~al\mbox{.}(2017)]%
        {pei2017deepXplore}
\bibfield{author}{\bibinfo{person}{Kexin Pei}, \bibinfo{person}{Yinzhi Cao}, \bibinfo{person}{Junfeng Yang}, {and} \bibinfo{person}{Suman Jana}.} \bibinfo{year}{2017}\natexlab{}.
\newblock \showarticletitle{DeepXplore: Automated Whitebox Testing of Deep Learning Systems}. In \bibinfo{booktitle}{\emph{Proceedings of the 26th Symposium on Operating Systems Principles, Shanghai, China, October 28-31, 2017}}. \bibinfo{publisher}{{ACM}}, \bibinfo{pages}{1--18}.
\newblock
\urldef\tempurl%
\url{https://doi.org/10.1145/3132747.3132785}
\showDOI{\tempurl}


\bibitem[Sengupta et~al\mbox{.}(2006)]%
        {sengupta2006research}
\bibfield{author}{\bibinfo{person}{Bikram Sengupta}, \bibinfo{person}{Satish Chandra}, {and} \bibinfo{person}{Vibha Sinha}.} \bibinfo{year}{2006}\natexlab{}.
\newblock \showarticletitle{A research agenda for distributed software development}. In \bibinfo{booktitle}{\emph{Proceedings of the 28th international conference on Software engineering}}. \bibinfo{pages}{731--740}.
\newblock


\bibitem[Sharif et~al\mbox{.}(2021)]%
        {Sharif2021DeepOrder}
\bibfield{author}{\bibinfo{person}{Aizaz Sharif}, \bibinfo{person}{Dusica Marijan}, {and} \bibinfo{person}{Marius Liaaen}.} \bibinfo{year}{2021}\natexlab{}.
\newblock \showarticletitle{DeepOrder: Deep Learning for Test Case Prioritization in Continuous Integration Testing}. In \bibinfo{booktitle}{\emph{{IEEE} International Conference on Software Maintenance and Evolution, {ICSME} 2021, Luxembourg, September 27 - October 1, 2021}}. \bibinfo{publisher}{{IEEE}}, \bibinfo{pages}{525--534}.
\newblock
\urldef\tempurl%
\url{https://doi.org/10.1109/ICSME52107.2021.00053}
\showDOI{\tempurl}


\bibitem[Shen et~al\mbox{.}(2024)]%
        {shen2024taskbenchbenchmarkinglargelanguage}
\bibfield{author}{\bibinfo{person}{Yongliang Shen}, \bibinfo{person}{Kaitao Song}, \bibinfo{person}{Xu Tan}, \bibinfo{person}{Wenqi Zhang}, \bibinfo{person}{Kan Ren}, \bibinfo{person}{Siyu Yuan}, \bibinfo{person}{Weiming Lu}, \bibinfo{person}{Dongsheng Li}, {and} \bibinfo{person}{Yueting Zhuang}.} \bibinfo{year}{2024}\natexlab{}.
\newblock \bibinfo{title}{TaskBench: Benchmarking Large Language Models for Task Automation}.
\newblock
\newblock
\showeprint[arxiv, accepted to neurips 2024]{2311.18760}~[cs.CL]
\urldef\tempurl%
\url{https://arxiv.org/abs/2311.18760}
\showURL{%
\tempurl}


\bibitem[Siddiq et~al\mbox{.}(2023)]%
        {66generatingUnitTests}
\bibfield{author}{\bibinfo{person}{Mohammed~Latif Siddiq}, \bibinfo{person}{Joanna Santos}, \bibinfo{person}{Ridwanul~Hasan Tanvir}, \bibinfo{person}{Noshin Ulfat}, \bibinfo{person}{Fahmid~Al Rifat}, {and} \bibinfo{person}{Vinicius~Carvalho Lopes}.} \bibinfo{year}{2023}\natexlab{}.
\newblock \showarticletitle{Exploring the Effectiveness of Large Language Models in Generating Unit Tests}.
\newblock \bibinfo{journal}{\emph{arXiv preprint arXiv:2305.00418}} (\bibinfo{year}{2023}).
\newblock


\bibitem[Spieker et~al\mbox{.}(2017)]%
        {Spieker2017Reinforcement}
\bibfield{author}{\bibinfo{person}{Helge Spieker}, \bibinfo{person}{Arnaud Gotlieb}, \bibinfo{person}{Dusica Marijan}, {and} \bibinfo{person}{Morten Mossige}.} \bibinfo{year}{2017}\natexlab{}.
\newblock \showarticletitle{Reinforcement learning for automatic test case prioritization and selection in continuous integration}. In \bibinfo{booktitle}{\emph{Proceedings of the 26th {ACM} {SIGSOFT} International Symposium on Software Testing and Analysis, Santa Barbara, CA, USA, July 10 - 14, 2017}}, \bibfield{editor}{\bibinfo{person}{Tevfik Bultan} {and} \bibinfo{person}{Koushik Sen}} (Eds.). \bibinfo{publisher}{{ACM}}, \bibinfo{pages}{12--22}.
\newblock
\urldef\tempurl%
\url{https://doi.org/10.1145/3092703.3092709}
\showDOI{\tempurl}


\bibitem[Sun et~al\mbox{.}(2010)]%
        {sun2010adiscriminative}
\bibfield{author}{\bibinfo{person}{Chengnian Sun}, \bibinfo{person}{David Lo}, \bibinfo{person}{Xiaoyin Wang}, \bibinfo{person}{Jing Jiang}, {and} \bibinfo{person}{Siau{-}Cheng Khoo}.} \bibinfo{year}{2010}\natexlab{}.
\newblock \showarticletitle{A discriminative model approach for accurate duplicate bug report retrieval}. In \bibinfo{booktitle}{\emph{Proceedings of the 32nd {ACM/IEEE} International Conference on Software Engineering - Volume 1, {ICSE} 2010, Cape Town, South Africa, 1-8 May 2010}}, \bibfield{editor}{\bibinfo{person}{Jeff Kramer}, \bibinfo{person}{Judith Bishop}, \bibinfo{person}{Premkumar~T. Devanbu}, {and} \bibinfo{person}{Sebasti{\'{a}}n Uchitel}} (Eds.). \bibinfo{publisher}{{ACM}}, \bibinfo{pages}{45--54}.
\newblock
\urldef\tempurl%
\url{https://doi.org/10.1145/1806799.1806811}
\showDOI{\tempurl}


\bibitem[Sun et~al\mbox{.}(2023)]%
        {204SmtSolverValidation}
\bibfield{author}{\bibinfo{person}{Maolin Sun}, \bibinfo{person}{Yibiao Yang}, \bibinfo{person}{Yang Wang}, \bibinfo{person}{Ming Wen}, \bibinfo{person}{Haoxiang Jia}, {and} \bibinfo{person}{Yuming Zhou}.} \bibinfo{year}{2023}\natexlab{}.
\newblock \showarticletitle{{SMT} Solver Validation Empowered by Large Pre-Trained Language Models}. In \bibinfo{booktitle}{\emph{38th {IEEE/ACM} International Conference on Automated Software Engineering, {ASE} 2023, Luxembourg, September 11-15, 2023}}. \bibinfo{publisher}{{IEEE}}, \bibinfo{pages}{1288--1300}.
\newblock
\urldef\tempurl%
\url{https://doi.org/10.1109/ASE56229.2023.00180}
\showDOI{\tempurl}


\bibitem[Sun et~al\mbox{.}(2020)]%
        {sun2020automatic}
\bibfield{author}{\bibinfo{person}{Zeyu Sun}, \bibinfo{person}{Jie~M. Zhang}, \bibinfo{person}{Mark Harman}, \bibinfo{person}{Mike Papadakis}, {and} \bibinfo{person}{Lu Zhang}.} \bibinfo{year}{2020}\natexlab{}.
\newblock \showarticletitle{Automatic testing and improvement of machine translation}. In \bibinfo{booktitle}{\emph{{ICSE} '20: 42nd International Conference on Software Engineering, Seoul, South Korea, 27 June - 19 July, 2020}}, \bibfield{editor}{\bibinfo{person}{Gregg Rothermel} {and} \bibinfo{person}{Doo{-}Hwan Bae}} (Eds.). \bibinfo{publisher}{{ACM}}, \bibinfo{pages}{974--985}.
\newblock
\urldef\tempurl%
\url{https://doi.org/10.1145/3377811.3380420}
\showDOI{\tempurl}


\bibitem[Tamrawi et~al\mbox{.}(2011)]%
        {tamrawi2011fuzzy}
\bibfield{author}{\bibinfo{person}{Ahmed Tamrawi}, \bibinfo{person}{Tung~Thanh Nguyen}, \bibinfo{person}{Jafar~M Al-Kofahi}, {and} \bibinfo{person}{Tien~N Nguyen}.} \bibinfo{year}{2011}\natexlab{}.
\newblock \showarticletitle{Fuzzy set and cache-based approach for bug triaging}. In \bibinfo{booktitle}{\emph{Proceedings of the 19th ACM SIGSOFT symposium and the 13th European conference on Foundations of software engineering}}. \bibinfo{pages}{365--375}.
\newblock


\bibitem[Tian et~al\mbox{.}(2018)]%
        {tian2018deeptest}
\bibfield{author}{\bibinfo{person}{Yuchi Tian}, \bibinfo{person}{Kexin Pei}, \bibinfo{person}{Suman Jana}, {and} \bibinfo{person}{Baishakhi Ray}.} \bibinfo{year}{2018}\natexlab{}.
\newblock \showarticletitle{DeepTest: automated testing of deep-neural-network-driven autonomous cars}. In \bibinfo{booktitle}{\emph{Proceedings of the 40th International Conference on Software Engineering, {ICSE} 2018, Gothenburg, Sweden, May 27 - June 03, 2018}}, \bibfield{editor}{\bibinfo{person}{Michel Chaudron}, \bibinfo{person}{Ivica Crnkovic}, \bibinfo{person}{Marsha Chechik}, {and} \bibinfo{person}{Mark Harman}} (Eds.). \bibinfo{publisher}{{ACM}}, \bibinfo{pages}{303--314}.
\newblock
\urldef\tempurl%
\url{https://doi.org/10.1145/3180155.3180220}
\showDOI{\tempurl}


\bibitem[Wang et~al\mbox{.}(2024a)]%
        {wang2023software}
\bibfield{author}{\bibinfo{person}{Junjie Wang}, \bibinfo{person}{Yuchao Huang}, \bibinfo{person}{Chunyang Chen}, \bibinfo{person}{Zhe Liu}, \bibinfo{person}{Song Wang}, {and} \bibinfo{person}{Qing Wang}.} \bibinfo{year}{2024}\natexlab{a}.
\newblock \showarticletitle{Software Testing with Large Language Model: Survey, Landscape, and Vision}.
\newblock \bibinfo{journal}{\emph{IEEE Transactions on Software Engineering}} (\bibinfo{year}{2024}).
\newblock


\bibitem[Wang et~al\mbox{.}(2021)]%
        {wang2021characterizing}
\bibfield{author}{\bibinfo{person}{Junjie Wang}, \bibinfo{person}{Song Wang}, \bibinfo{person}{Jianfeng Chen}, \bibinfo{person}{Tim Menzies}, \bibinfo{person}{Qiang Cui}, \bibinfo{person}{Miao Xie}, {and} \bibinfo{person}{Qing Wang}.} \bibinfo{year}{2021}\natexlab{}.
\newblock \showarticletitle{Characterizing Crowds to Better Optimize Worker Recommendation in Crowdsourced Testing}.
\newblock \bibinfo{journal}{\emph{{IEEE} Trans. Software Eng.}} \bibinfo{volume}{47}, \bibinfo{number}{6} (\bibinfo{year}{2021}), \bibinfo{pages}{1259--1276}.
\newblock
\urldef\tempurl%
\url{https://doi.org/10.1109/TSE.2019.2918520}
\showDOI{\tempurl}


\bibitem[Wang et~al\mbox{.}(2019)]%
        {wang2019isense}
\bibfield{author}{\bibinfo{person}{Junjie Wang}, \bibinfo{person}{Ye Yang}, \bibinfo{person}{Rahul Krishna}, \bibinfo{person}{Tim Menzies}, {and} \bibinfo{person}{Qing Wang}.} \bibinfo{year}{2019}\natexlab{}.
\newblock \showarticletitle{iSENSE: Completion-aware crowdtesting management}. In \bibinfo{booktitle}{\emph{2019 IEEE/ACM 41st international conference on software engineering (ICSE)}}. IEEE, \bibinfo{pages}{912--923}.
\newblock


\bibitem[Wang et~al\mbox{.}(2020)]%
        {wang2020contextaware}
\bibfield{author}{\bibinfo{person}{Junjie Wang}, \bibinfo{person}{Ye Yang}, \bibinfo{person}{Song Wang}, \bibinfo{person}{Yuanzhe Hu}, \bibinfo{person}{Dandan Wang}, {and} \bibinfo{person}{Qing Wang}.} \bibinfo{year}{2020}\natexlab{}.
\newblock \showarticletitle{Context-aware in-process crowdworker recommendation}. In \bibinfo{booktitle}{\emph{{ICSE} '20: 42nd International Conference on Software Engineering, Seoul, South Korea, 27 June - 19 July, 2020}}, \bibfield{editor}{\bibinfo{person}{Gregg Rothermel} {and} \bibinfo{person}{Doo{-}Hwan Bae}} (Eds.). \bibinfo{publisher}{{ACM}}, \bibinfo{pages}{1535--1546}.
\newblock
\urldef\tempurl%
\url{https://doi.org/10.1145/3377811.3380380}
\showDOI{\tempurl}


\bibitem[Wang et~al\mbox{.}(2024b)]%
        {wang2024pandalmautomaticevaluationbenchmark}
\bibfield{author}{\bibinfo{person}{Yidong Wang}, \bibinfo{person}{Zhuohao Yu}, \bibinfo{person}{Zhengran Zeng}, \bibinfo{person}{Linyi Yang}, \bibinfo{person}{Cunxiang Wang}, \bibinfo{person}{Hao Chen}, \bibinfo{person}{Chaoya Jiang}, \bibinfo{person}{Rui Xie}, \bibinfo{person}{Jindong Wang}, \bibinfo{person}{Xing Xie}, \bibinfo{person}{Wei Ye}, \bibinfo{person}{Shikun Zhang}, {and} \bibinfo{person}{Yue Zhang}.} \bibinfo{year}{2024}\natexlab{b}.
\newblock \bibinfo{title}{PandaLM: An Automatic Evaluation Benchmark for LLM Instruction Tuning Optimization}.
\newblock
\newblock
\showeprint[arxiv, accepted by iclr 2024]{2306.05087}~[cs.CL]
\urldef\tempurl%
\url{https://arxiv.org/abs/2306.05087}
\showURL{%
\tempurl}


\bibitem[Wei et~al\mbox{.}(2022)]%
        {wei2022freelunch}
\bibfield{author}{\bibinfo{person}{Anjiang Wei}, \bibinfo{person}{Yinlin Deng}, \bibinfo{person}{Chenyuan Yang}, {and} \bibinfo{person}{Lingming Zhang}.} \bibinfo{year}{2022}\natexlab{}.
\newblock \showarticletitle{Free lunch for testing: fuzzing deep-learning libraries from open source}. In \bibinfo{booktitle}{\emph{Proceedings of the 44th International Conference on Software Engineering}} (Pittsburgh, Pennsylvania) \emph{(\bibinfo{series}{ICSE '22})}. \bibinfo{publisher}{Association for Computing Machinery}, \bibinfo{address}{New York, NY, USA}, \bibinfo{pages}{995–1007}.
\newblock
\showISBNx{9781450392211}
\urldef\tempurl%
\url{https://doi.org/10.1145/3510003.3510041}
\showDOI{\tempurl}


\bibitem[Xi et~al\mbox{.}(2023)]%
        {xi2023rise}
\bibfield{author}{\bibinfo{person}{Zhiheng Xi}, \bibinfo{person}{Wenxiang Chen}, \bibinfo{person}{Xin Guo}, \bibinfo{person}{Wei He}, \bibinfo{person}{Yiwen Ding}, \bibinfo{person}{Boyang Hong}, \bibinfo{person}{Ming Zhang}, \bibinfo{person}{Junzhe Wang}, \bibinfo{person}{Senjie Jin}, \bibinfo{person}{Enyu Zhou}, \bibinfo{person}{Rui Zheng}, \bibinfo{person}{Xiaoran Fan}, \bibinfo{person}{Xiao Wang}, \bibinfo{person}{Limao Xiong}, \bibinfo{person}{Yuhao Zhou}, \bibinfo{person}{Weiran Wang}, \bibinfo{person}{Changhao Jiang}, \bibinfo{person}{Yicheng Zou}, \bibinfo{person}{Xiangyang Liu}, \bibinfo{person}{Zhangyue Yin}, \bibinfo{person}{Shihan Dou}, \bibinfo{person}{Rongxiang Weng}, \bibinfo{person}{Wensen Cheng}, \bibinfo{person}{Qi Zhang}, \bibinfo{person}{Wenjuan Qin}, \bibinfo{person}{Yongyan Zheng}, \bibinfo{person}{Xipeng Qiu}, \bibinfo{person}{Xuanjing Huang}, {and} \bibinfo{person}{Tao Gui}.} \bibinfo{year}{2023}\natexlab{}.
\newblock \bibinfo{title}{The Rise and Potential of Large Language Model Based Agents: A Survey}.
\newblock
\newblock
\showeprint[arxiv]{2309.07864}~[cs.AI]


\bibitem[Xia et~al\mbox{.}(2017)]%
        {xia2017improving}
\bibfield{author}{\bibinfo{person}{Xin Xia}, \bibinfo{person}{David Lo}, \bibinfo{person}{Ying Ding}, \bibinfo{person}{Jafar~M. Al{-}Kofahi}, \bibinfo{person}{Tien~N. Nguyen}, {and} \bibinfo{person}{Xinyu Wang}.} \bibinfo{year}{2017}\natexlab{}.
\newblock \showarticletitle{Improving Automated Bug Triaging with Specialized Topic Model}.
\newblock \bibinfo{journal}{\emph{{IEEE} Trans. Software Eng.}} \bibinfo{volume}{43}, \bibinfo{number}{3} (\bibinfo{year}{2017}), \bibinfo{pages}{272--297}.
\newblock
\urldef\tempurl%
\url{https://doi.org/10.1109/TSE.2016.2576454}
\showDOI{\tempurl}


\bibitem[Yang et~al\mbox{.}(2016)]%
        {yang2016combining}
\bibfield{author}{\bibinfo{person}{Xinli Yang}, \bibinfo{person}{David Lo}, \bibinfo{person}{Xin Xia}, \bibinfo{person}{Lingfeng Bao}, {and} \bibinfo{person}{Jianling Sun}.} \bibinfo{year}{2016}\natexlab{}.
\newblock \showarticletitle{Combining Word Embedding with Information Retrieval to Recommend Similar Bug Reports}. In \bibinfo{booktitle}{\emph{27th {IEEE} International Symposium on Software Reliability Engineering, {ISSRE} 2016, Ottawa, ON, Canada, October 23-27, 2016}}. \bibinfo{publisher}{{IEEE} Computer Society}, \bibinfo{pages}{127--137}.
\newblock
\urldef\tempurl%
\url{https://doi.org/10.1109/ISSRE.2016.33}
\showDOI{\tempurl}


\bibitem[Yang et~al\mbox{.}(2022)]%
        {yang2022survey}
\bibfield{author}{\bibinfo{person}{Yanming Yang}, \bibinfo{person}{Xin Xia}, \bibinfo{person}{David Lo}, {and} \bibinfo{person}{John Grundy}.} \bibinfo{year}{2022}\natexlab{}.
\newblock \showarticletitle{A survey on deep learning for software engineering}.
\newblock \bibinfo{journal}{\emph{ACM Computing Surveys (CSUR)}} \bibinfo{volume}{54}, \bibinfo{number}{10s} (\bibinfo{year}{2022}), \bibinfo{pages}{1--73}.
\newblock


\bibitem[Yang et~al\mbox{.}(2024)]%
        {yang2024stealthy}
\bibfield{author}{\bibinfo{person}{Zhou Yang}, \bibinfo{person}{Bowen Xu}, \bibinfo{person}{Jie~M Zhang}, \bibinfo{person}{Hong~Jin Kang}, \bibinfo{person}{Jieke Shi}, \bibinfo{person}{Junda He}, {and} \bibinfo{person}{David Lo}.} \bibinfo{year}{2024}\natexlab{}.
\newblock \showarticletitle{Stealthy backdoor attack for code models}.
\newblock \bibinfo{journal}{\emph{IEEE Transactions on Software Engineering}} (\bibinfo{year}{2024}).
\newblock


\bibitem[Yaraghi et~al\mbox{.}(2022)]%
        {yaraghi2022scalable}
\bibfield{author}{\bibinfo{person}{Ahmadreza~Saboor Yaraghi}, \bibinfo{person}{Mojtaba Bagherzadeh}, \bibinfo{person}{Nafiseh Kahani}, {and} \bibinfo{person}{Lionel~C Briand}.} \bibinfo{year}{2022}\natexlab{}.
\newblock \showarticletitle{Scalable and accurate test case prioritization in continuous integration contexts}.
\newblock \bibinfo{journal}{\emph{IEEE Transactions on Software Engineering}} \bibinfo{volume}{49}, \bibinfo{number}{4} (\bibinfo{year}{2022}), \bibinfo{pages}{1615--1639}.
\newblock


\bibitem[Yuan et~al\mbox{.}(2023)]%
        {64noUnitTest}
\bibfield{author}{\bibinfo{person}{Zhiqiang Yuan}, \bibinfo{person}{Yiling Lou}, \bibinfo{person}{Mingwei Liu}, \bibinfo{person}{Shiji Ding}, \bibinfo{person}{Kaixin Wang}, \bibinfo{person}{Yixuan Chen}, {and} \bibinfo{person}{Xin Peng}.} \bibinfo{year}{2023}\natexlab{}.
\newblock \showarticletitle{No More Manual Tests? Evaluating and Improving ChatGPT for Unit Test Generation}.
\newblock \bibinfo{journal}{\emph{arXiv preprint arXiv:2305.04207}} (\bibinfo{year}{2023}).
\newblock


\bibitem[Zamprogno et~al\mbox{.}(2022)]%
        {zamprogno2022dynamic}
\bibfield{author}{\bibinfo{person}{Lucas Zamprogno}, \bibinfo{person}{Braxton Hall}, \bibinfo{person}{Reid Holmes}, {and} \bibinfo{person}{Joanne~M Atlee}.} \bibinfo{year}{2022}\natexlab{}.
\newblock \showarticletitle{Dynamic human-in-the-loop assertion generation}.
\newblock \bibinfo{journal}{\emph{IEEE Transactions on Software Engineering}} \bibinfo{volume}{49}, \bibinfo{number}{4} (\bibinfo{year}{2022}), \bibinfo{pages}{2337--2351}.
\newblock


\bibitem[Zhang et~al\mbox{.}(2019)]%
        {zhang2019predictive}
\bibfield{author}{\bibinfo{person}{Jie Zhang}, \bibinfo{person}{Lingming Zhang}, \bibinfo{person}{Mark Harman}, \bibinfo{person}{Dan Hao}, \bibinfo{person}{Yue Jia}, {and} \bibinfo{person}{Lu Zhang}.} \bibinfo{year}{2019}\natexlab{}.
\newblock \showarticletitle{Predictive Mutation Testing}.
\newblock \bibinfo{journal}{\emph{{IEEE} Trans. Software Eng.}} \bibinfo{volume}{45}, \bibinfo{number}{9} (\bibinfo{year}{2019}), \bibinfo{pages}{898--918}.
\newblock
\urldef\tempurl%
\url{https://doi.org/10.1109/TSE.2018.2809496}
\showDOI{\tempurl}


\bibitem[Zhang et~al\mbox{.}(2022a)]%
        {Zhang2022Machine}
\bibfield{author}{\bibinfo{person}{Jie~M. Zhang}, \bibinfo{person}{Mark Harman}, \bibinfo{person}{Lei Ma}, {and} \bibinfo{person}{Yang Liu}.} \bibinfo{year}{2022}\natexlab{a}.
\newblock \showarticletitle{Machine Learning Testing: Survey, Landscapes and Horizons}.
\newblock \bibinfo{journal}{\emph{{IEEE} Trans. Software Eng.}} \bibinfo{volume}{48}, \bibinfo{number}{2} (\bibinfo{year}{2022}), \bibinfo{pages}{1--36}.
\newblock


\bibitem[Zhang et~al\mbox{.}(2022b)]%
        {zhang2022machineLearningTesting}
\bibfield{author}{\bibinfo{person}{Jie~M. Zhang}, \bibinfo{person}{Mark Harman}, \bibinfo{person}{Lei Ma}, {and} \bibinfo{person}{Yang Liu}.} \bibinfo{year}{2022}\natexlab{b}.
\newblock \showarticletitle{Machine Learning Testing: Survey, Landscapes and Horizons}.
\newblock \bibinfo{journal}{\emph{{IEEE} Trans. Software Eng.}} \bibinfo{volume}{48}, \bibinfo{number}{2} (\bibinfo{year}{2022}), \bibinfo{pages}{1--36}.
\newblock


\bibitem[Zhang et~al\mbox{.}(2018)]%
        {ZhangZZ0K2018Deeproad}
\bibfield{author}{\bibinfo{person}{Mengshi Zhang}, \bibinfo{person}{Yuqun Zhang}, \bibinfo{person}{Lingming Zhang}, \bibinfo{person}{Cong Liu}, {and} \bibinfo{person}{Sarfraz Khurshid}.} \bibinfo{year}{2018}\natexlab{}.
\newblock \showarticletitle{DeepRoad: GAN-based metamorphic testing and input validation framework for autonomous driving systems}. In \bibinfo{booktitle}{\emph{Proceedings of the 33rd {ACM/IEEE} International Conference on Automated Software Engineering, {ASE} 2018, Montpellier, France, September 3-7, 2018}}, \bibfield{editor}{\bibinfo{person}{Marianne Huchard}, \bibinfo{person}{Christian K{\"{a}}stner}, {and} \bibinfo{person}{Gordon Fraser}} (Eds.). \bibinfo{publisher}{{ACM}}, \bibinfo{pages}{132--142}.
\newblock
\urldef\tempurl%
\url{https://doi.org/10.1145/3238147.3238187}
\showDOI{\tempurl}


\bibitem[Zhang et~al\mbox{.}(2023)]%
        {zhang2023duplicate}
\bibfield{author}{\bibinfo{person}{Ting Zhang}, \bibinfo{person}{DongGyun Han}, \bibinfo{person}{Venkatesh Vinayakarao}, \bibinfo{person}{Ivana~Clairine Irsan}, \bibinfo{person}{Bowen Xu}, \bibinfo{person}{Ferdian Thung}, \bibinfo{person}{David Lo}, {and} \bibinfo{person}{Lingxiao Jiang}.} \bibinfo{year}{2023}\natexlab{}.
\newblock \showarticletitle{Duplicate Bug Report Detection: How Far Are We?}
\newblock \bibinfo{journal}{\emph{{ACM} Trans. Softw. Eng. Methodol.}} \bibinfo{volume}{32}, \bibinfo{number}{4} (\bibinfo{year}{2023}), \bibinfo{pages}{97:1--97:32}.
\newblock
\urldef\tempurl%
\url{https://doi.org/10.1145/3576042}
\showDOI{\tempurl}


\bibitem[Zhou et~al\mbox{.}(2020)]%
        {zhou2020Deepbillboard}
\bibfield{author}{\bibinfo{person}{Husheng Zhou}, \bibinfo{person}{Wei Li}, \bibinfo{person}{Zelun Kong}, \bibinfo{person}{Junfeng Guo}, \bibinfo{person}{Yuqun Zhang}, \bibinfo{person}{Bei Yu}, \bibinfo{person}{Lingming Zhang}, {and} \bibinfo{person}{Cong Liu}.} \bibinfo{year}{2020}\natexlab{}.
\newblock \showarticletitle{DeepBillboard: systematic physical-world testing of autonomous driving systems}. In \bibinfo{booktitle}{\emph{{ICSE} '20: 42nd International Conference on Software Engineering, Seoul, South Korea, 27 June - 19 July, 2020}}, \bibfield{editor}{\bibinfo{person}{Gregg Rothermel} {and} \bibinfo{person}{Doo{-}Hwan Bae}} (Eds.). \bibinfo{publisher}{{ACM}}, \bibinfo{pages}{347--358}.
\newblock
\urldef\tempurl%
\url{https://doi.org/10.1145/3377811.3380422}
\showDOI{\tempurl}


\end{thebibliography}


\end{document}